\documentclass[fleqn,10pt]{wlscirep}
\usepackage[utf8]{inputenc}
\usepackage[T1]{fontenc}
\usepackage{lineno}

\usepackage{color}

\graphicspath{ {pics/} }
\title{High Transmission in 120-degree Sharp Bends of Inversion-symmetric and Inversion-asymmetric Photonic Crystal Waveguides}

\author[1,2]{Wei Dai}
\author[1,2]{Taiki Yoda}
\author[1]{Yuto Moritake}
\author[2,3]{Masaaki Ono}
\author[2,3]{Eiichi Kuramochi}
\author[1,2,3,*]{Masaya Notomi}
\affil[1]{Department of Physics, Tokyo Institute of Technology, 2-12-1 Ookayama, Meguro-ku, Tokyo 152-8550, Japan}
\affil[2]{NTT Basic Research Laboratories, NTT Corporation, 3-1 Moriosato-Wakamiya, Atsugi, 243-0198, Japan}
\affil[3]{Nanophotonics Center, NTT Corporation, 3-1 Moriosato-Wakamiya,
Atsugi, 243-0198, Japan}
\affil[*]{E-mail: notomi@phys.titech.ac.jp}

\begin{abstract}
Bending loss is one of the serious problems for constructing nanophotonic integrated circuits. Recently, many works
reported that valley photonic crystals (VPhCs) enable significantly high transmission via 120-degree sharp bends. 
However, it is unclear whether
the high bend-transmission results directly from the valley-photonic effects, which are based on the breaking of inversion symmetry. In this study, we conduct a series of comparative
numerical and experimental investigations of bend-transmission in various triangular PhCs with and without inversion symmetry
and reveal that the high bend-transmission is solely determined by the domain-wall configuration and independent of the existence
of the inversion symmetry. Preliminary analysis of the polarization distribution indicates that high bend-transmissions are closely related to the appearance of local topological polarization singularities near the bending section. Our work demonstrates that high transmission can be achieved in a much wider family of PhC waveguides, which may provide novel designs for low-loss nanophotonic integrated circuits with enhanced flexibility and a new understanding of the nature of valley-photonics

\end{abstract}

\begin{document}

\flushbottom
\maketitle

\section*{Introduction}
 Photonic crystal waveguides (PhCWGs) that support highly confined light modes have wide applications in telecommunication and data processing\cite{ref1,ref2,ref3,ref4}. Recently, valley photonic crystals (VPhCs), an optic implementation of valley Hall effect\cite{ref5_background,ref6_background,ref7_background}, have offered large-scale, all-dielectric designs for PhCWGs. The heart of valley-photonic properties is the breaking of inversion symmetry, which gives rise to non-trivial Berry curvatures around the \(K\) and \(K’\) points, leading to a distinct valley Chern number and topological bandgaps, and suggesting the possibility of suppressed backscattering\cite{Mref22_Dongjianwen2017_HCbulk,Mref17_Yang2018_h23,Mref16_Ma2016_s0}. When two VPhCs with different Chern numbers are connected by an interface, topological domain-wall modes appear within the bandgaps. 

Bending loss is one of the most serious problems in constructing photonic integrated circuits employing nanophotonics\cite{Bref_1,Bref2,Bref3,Bref4}. This is because sharp bends exhibit significantly large reflections when the bending radius is comparable to the wavelength of light. For example, a simple single-missing-hole line defect waveguide (so-called, W1), the most widely-used PhC waveguides, have large reflections at 120-degree bends unless sophisticatedly modified at the corners\cite{Bref_1,Bref2,Bref3}. 
In contrast, many recent reports showed that various VPhC waveguides exhibit extraordinarily high transmission through 120-degree bends within a wide frequency range \cite{Mref1_Majiawen2019_h23,Mref2_Yamaguchi_2019_h23,Mref3_Mikhail2019_h23,Mref4_ChenXD2017_h23,Mref4.1_HeXT2022_t13,Mref26_h13_Kumar2022,Mref27_H13_Arora2021,Mref5_HAN2021_t16,Mref7_He2019_h16,Mref8_JalaliMehrabad2020_h56,Mref9_Yoshimi:21_h56,Mref10_Yoshimi:20_h56,Mref12_Gao2017_h56,Mref11_Cehn2019_h16,Mref13_Wu2017_s0,Mref14_Zhang2020_s0,Mref15_Kang2018_s0,Mref19_Zeng2020_s0,Mref20_DU2020_s0,Mref24_Wang_2019_S0,Mref23_XiangXi_HCzigzag}. This interesting property of VPhCs has attracted considerable attention. Since backscattering suppression is generally expected for edge or domain-wall modes in topological insulators, reflection-free transmission has been considered a topological feature of VPhCs. Naively, if one assumes that inter-valley scattering is prohibited, valley spins should be conserved, and thus the back reflection should be suppressed.          

However, no unambiguous demonstration proves the direct relationship between high transmission in bends and the topological properties. In fact, we believe some ambiguities remain. (1) Theoretically, it is not apparent whether inter-valley scattering could be prohibited or valley spins could be conserved at bends. In the usual situation, valley spin can be easily flipped upon reflection. For example, Arregui et al.\cite{Mref25_Guillermo_disorderH23} numerically showed that the suppression of backscattering occurs only at ultraslow light modes in straight VPhCWGs with minimal disorders, and a recent experimental work supported their conclusion\cite{disroder2_Rosiek2023}. This minimal perturbation condition is hardly satisfied in the transmission in 120-degree bends.  (2) Experimentally or numerically, it has not been directly proved that the high transmission is due to the valley-photonic effects. In some previous studies\cite{Mref10_Yoshimi:20_h56,Mref27_H13_Arora2021}, W1WGs are used as the reference to verify the high transmission in VPhCWGs. However, the domain-wall configurations of W1WGs and VPhCWGs are largely different. Besides the inversion symmetry in the bulk lattice, W1WGs also have larger waveguide widths and their domain-wall configuration is not compatible with a honeycomb structure. Therefore, there remain possibilities that the high transmissions result from the difference in the domain-wall configuration instead of the topological effect.

In this study, to identify the origin of high transmission in 120-degree bent PhCWGs,  we separate the effect of inversion symmetry and the domain-wall configuration by employing a specific model structure in which we can vary the interface condition and the inversion symmetry separately. We show theoretical treatments first, following extensive experimental works. Our theoretical and experimental studies reveal a surprising result in which the high bend-transmittance appears irrespective of the existence of the inversion symmetry. It is shown that the high bend-transmission can be realized in a much wider range of structures than previously expected. Since the breaking inversion symmetry is the origin of the valley photonic effect, our finding indicates that the high bend-transmittance does not originate from the valley effect. Furthermore, we carefully investigate the appearance condition of the high bent transmission for various interface structures, and finally give an intriguing insight into the origin of the high bend-transmission.

\section*{Results}

\subsection*{Model structures}

We propose to employ a systematic model representing various domain-wall configurations with and without the inversion symmetry. In a honeycomb lattice, restoring the inversion symmetry closes the bandgap and thus no domain-wall modes remain. Here, we adopt triangular-lattice air-hole PhCs as shown in the inset image of Fig.1(a,b). We manipulate the inversion symmetry by changing the hole shapes. The bulk lattices are either inversion-symmetric with circular air holes (IS-PhCs) or inversion-asymmetric with triangular air holes (IA-PhCs). 

\begin{figure}[hbt!]
\centering
\includegraphics[width=0.8\textwidth]{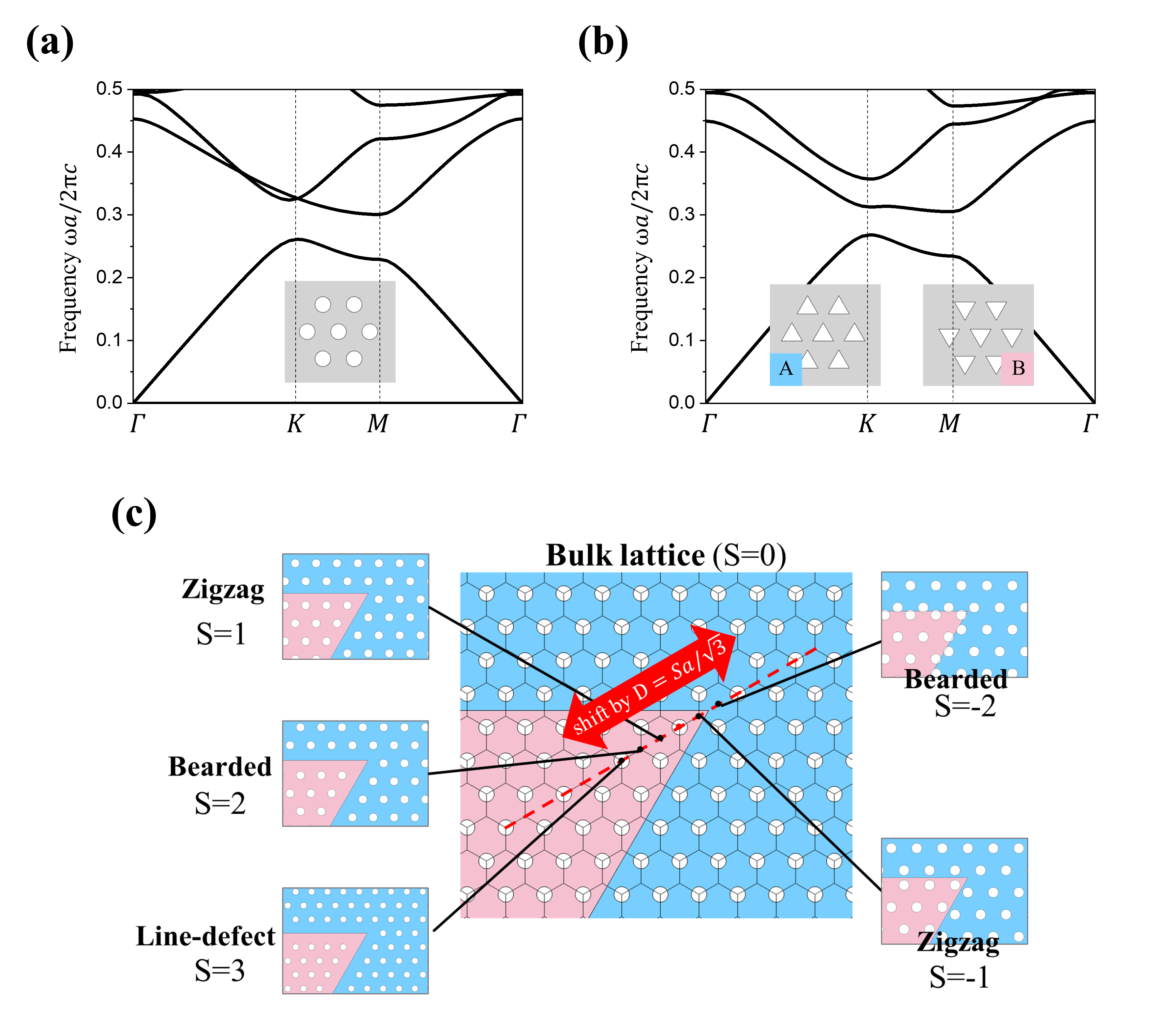}
\caption{The band structures of triangular lattice Si-slab PhCs (a) with and (b) without inversion symmetry. Inset in (a) shows the IS-PhC with circular air holes in the silicon slab. The lattice constant is 400 nm. The radius of air holes is 102 nm. Inset in (b) shows the A-type and B-type IA-PhC with triangular holes. The lattice constant is 400 nm. The side length of air holes is 277 nm. The effective refractive index of silicon is 2.65. (c) conceptual illustration of the universal design of triangular lattice waveguides that are compatible with 120-degree sharp bends. This large picture shows a bulk triangular lattice (\(S=0\)) with inversion symmetry (circular holes). The bent interface forms a 60-degree angle. The red dotted line is the angle bisector. The smaller pictures show the five domain-wall configurations obtained by shifting the red region when the shifting parameter \(S\) is -2,-1,1,2, and 3. }
\label{fig:figure1}
\end{figure}

IA-PhCWGs based on a triangular lattice have been already proposed and high transmissions through 120-degree bends have been observed\cite{Mref4.1_HeXT2022_t13,Mref5_HAN2021_t16,Yoda:19,Mref13_Wu2017_s0,Mref14_Zhang2020_s0,Mref19_Zeng2020_s0}.  Here, we employ slightly different IA-PhCWGs based on triangular air-hole lattices and focus on the first bandgap in TE-polarization (Fig.1(a,b)). This first bandgap is most widely used in PhC waveguides, including W1WGs. Since the Dirac degeneracy occurs between the second and third bands at \(K\) points, the first bandgap always exists even with the inversion symmetry. Therefore, one can easily alternate the inversion symmetry without affecting the lattice configuration. The crucial point is that the valley-photonic effect still exists in this case. Recently, we investigated\cite{Yoda:19,Yoda:unpublished} this type of triangular-hole PhCs without inversion symmetry (IA-PhCs, shown in Fig.1(b)) and found that they exhibit the valley-photonic effects in an essentially similar way to conventional VPhCs. We theoretically confirmed that these PhCs show large nontrivial Berry curvature around the \(K\) and \(K’\) points of the first and second bands. Interestingly, the Berry curvature in these bands is even larger than that of the third band, which originates from the Dirac point in IS-PhCs. Moreover, the first and second bands have opposite distinctive angular momentum. Thus, these valleys naturally lead to various valley-photonic effects.

Starting from this bulk design, a wide variety of domain-walls can be constructed by simply shifting the lattice in the half-space. As shown in Fig.1(c), we divide the triangular lattice PhC with a 60-degree angle boundary into the blue and red regions. We can shift the two divided regions along the angle bisector (dotted red line) to introduce a line defect to the bulk lattice and thus construct a 120-degree bent waveguide. Waveguides constructed in this manner can be characterized by the shift direction and shift distance \(D\). Here, we define a shifting parameter \(S=\sqrt{3}D/a\). \(S\) is positive (negative) when the red region is shifted away from (towards) the blue region. When \(S=3\), the waveguide is a conventional W1WG with circular holes. When \(S=\pm1\), the domain wall is a zigzag interface. When \(S=\pm2\), the domain wall is a bearded interface. Note that when \(S\) is even, the waveguide is mirror-symmetric. When \(S\) is odd, the waveguide is glide-symmetric. When \(S\) is a non-integer, the interface has neither mirror symmetry nor glide symmetry.

\textcolor{black}{We can also classify the domain-wall configuration of other types of VPhCs previously reported in a similar manner.}
%These domain wall configurations are essentially similar to those in previous VPhC studies.%
For example, the zigzag interface in reference [19]\cite{Mref4.1_HeXT2022_t13} corresponds to \(S=-1\), and the bearded interface in reference [22]\cite{Mref5_HAN2021_t16} corresponds to \(S=-2\). In addition, honeycomb lattice VPhCs can be classified in the same manner if we focus on the configuration of large holes (or pillars). When \(S=\pm1\), the lattice becomes one of the sublattices of a zigzag interface honeycomb lattice \cite{Mref1_Majiawen2019_h23,Mref2_Yamaguchi_2019_h23,Mref3_Mikhail2019_h23,Mref4_ChenXD2017_h23,Mref25_Guillermo_disorderH23,Mref26_h13_Kumar2022,Mref4.1_HeXT2022_t13}. When \(S=\pm2\), the lattice becomes one of the sublattices of a bearded interface honeycomb lattice\cite{Mref5_HAN2021_t16,Mref7_He2019_h16,Mref8_JalaliMehrabad2020_h56,Mref9_Yoshimi:21_h56,Mref10_Yoshimi:20_h56,Mref12_Gao2017_h56}. The sign change of \(S\) corresponds to the sublattice exchange in a honeycomb lattice.
We assume that the broken inversion symmetry and domain-wall types (parameter S) would both be able to affect the high transmission through sharp bends. The comparison between VPhCWGs and W1WGs alone cannot distinguish between the two factors. Therefore, in this study, we compare the light transmission through 120-degree sharp bends between IS-PhCWGs and IA-PhCWGs having the same domain-wall type and then examine different domain-wall types.  We mainly investigate the aforementioned five types of interfaces: \(S=-2,-1,1,2,3\).

\subsection*{Numerical studies of Z-shaped waveguides}

\subsubsection*{(i) W1WG (S = 3, mirror-symmetric waveguides with inversion symmetry)}

\begin{figure}[hbt!]
\includegraphics[width=1.0\textwidth]{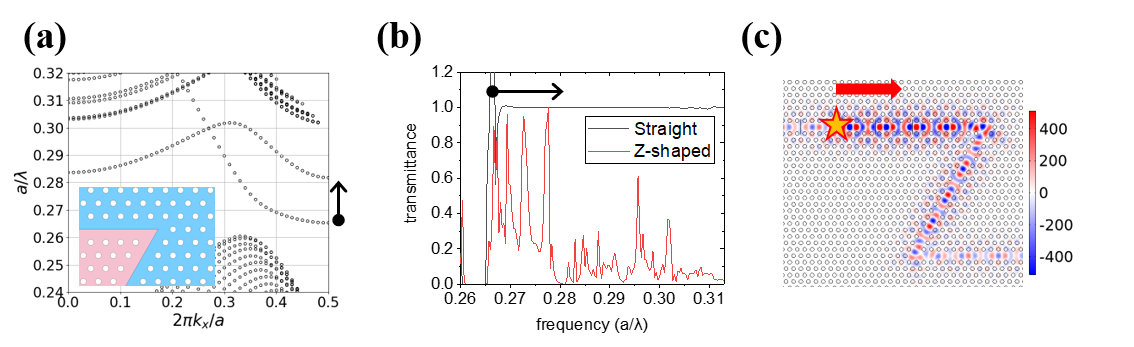}
\caption{Calculation results of the straight and Z-shaped \(S = 3\) IS-PhCWGs (W1WGs). (a) The 2-dimensional band structure. The black arrow shows the single-mode region. The inset shows one corner of the bent waveguide. (b) The transmittance spectra of the straight W1WG (black curve) and Z-shaped W1WG. The average transmittance is 0.47, and the F-P reflectivity is 0.40. (c) The out-of-plane magnetic field \(H_z\) of the Z-shaped W1WG at \(a/\lambda = 0.270\). The transmittance is 0.39. The five-pointed star indicates the location of the wave source. The arrow points toward the propagation direction. }
\label{fig:figure2}
\end{figure}

Here we numerically investigate Z-shaped waveguides in Si PhC slabs, consisting of a pair of 120-degree bends with the middle segment length of \(20a\). Firstly, we investigate the configuration of \(S=3\) with circular holes, corresponding to W1WG, which
is mirror-symmetric and possesses inversion symmetry. It has an even and an odd band in the PBG (Fig.2(a)), and here we focus on the even modes in the lower-frequency band. As shown in Fig.2(a) with a black arrow, there is a single-mode region in the frequency range \(a/\lambda= 0.265-0.282\). Within this single-mode region, we can see a clear transmission contrast between the straight (Fig.2(b), black curve) and the bent (Fig.2(b), red curve) waveguides. There are strong ripples in the spectrum of the bent waveguide. We calculate the corresponding cavity length to be around \(21.5a\) from the free spectral range (FSR) of the ripples. This length is very close to the middle segment’s length \(20a\) in the Z-shaped waveguide. Therefore, we confirm that these are Fabry-Pérot (F-P) ripples resulting from strong reflection at the two bends. In order to quantitatively analyze the transmittance, we calculate the average transmittance \(T_{av}\) and the F-P reflectivity (\(R_{FP}\)) in a single-mode region discarding the ultraslow-light region (see Method section for the definition of \(T_{av}\) and \(R_{FP}\)).  For the W1WG the frequency range is \(a/\lambda = 0.269-0.278\). The calculated  \(T_{av}\) and \(R_{FP}\) are 0.47 and 0.40.

As a typical example of the field distribution for the Z-shaped waveguide, \(H_z\) near a bend at \(a/\lambda = 0.270\) is shown in Fig.2(c) where the transmittance is 0.39. The five-pointed star indicates the location of the light source, exciting a rightward propagating mode. The light intensity is fairly reduced at the output port, and there is significant reflected intensity behind the excitation source. Hereafter, we investigate other types of waveguides focusing on their \(T_{av}\) and \(R_{FP}\).

\begin{figure}[hbt!]
\includegraphics[width=1.0\textwidth]{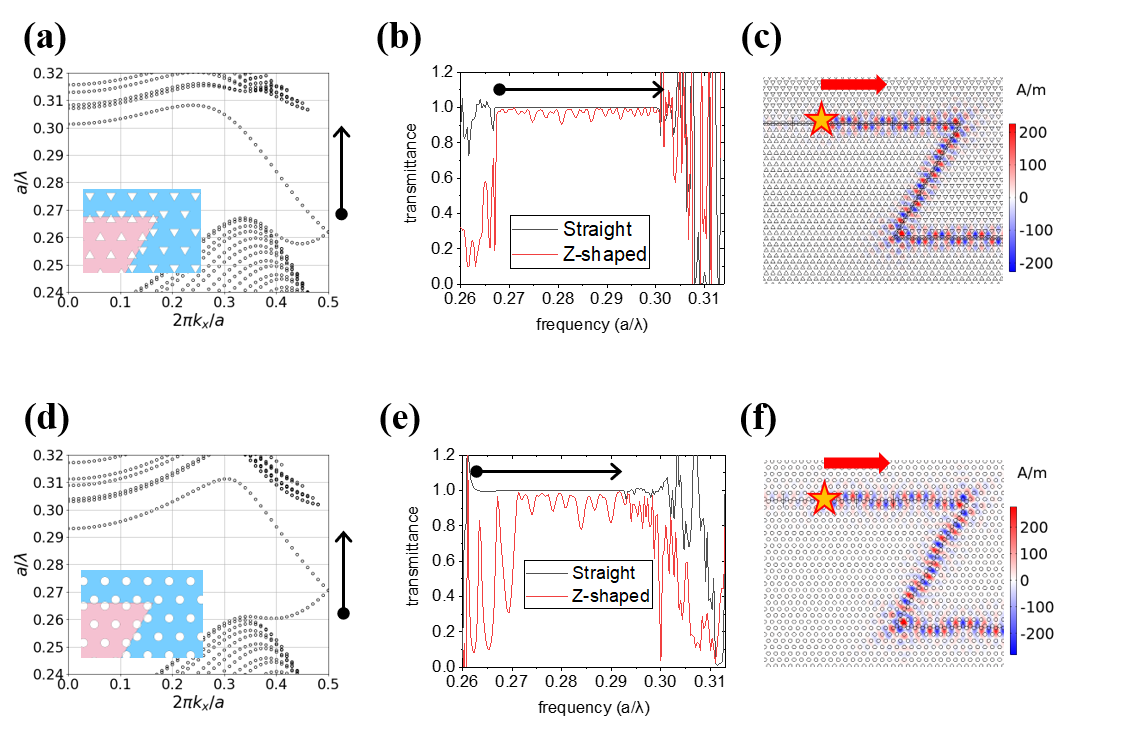}
\caption{Calculation results of the straight and Z-shaped \(S = -2\) PhCWGs. (a) The PBS of \(S=-2\) IA-PhCWG. The black arrow shows the single-mode region. The inset shows one corner of the bent waveguide. (b) The transmittance spectra of the straight \(S=-2\) IA-PhCWG (black curve) and Z-shaped \(S=-2\) IA-PhCWG (red curve). The average transmittance is 0.97. The F-P reflectivity is 0.03. (c) The out-of-plane magnetic field \(H_z\) of the Z-shaped \(S=-2\) IA-PhCWG at \(a/\lambda = 0.280\), with a transmittance of 0.97. The Hz has mixed spatial parity and meanders along the glide-symmetric interface. (d) The PBS of \(S=-2\) IS-PhCWG. The black arrow shows the single-mode region. The inset shows one corner of the bent waveguide. (e) The transmittance spectra of the straight \(S=-2\) IS-PhCWG (black curve) and Z-shaped \(S=-2\) IS-PhCWG (red curve). The average transmittance of the upper band is 0.94. The F-P reflectivity of the upper band is 0.06. The average transmittance of the lower band is 0.40. The F-P reflectivity of the lower band is 0.52. (f) The out-of-plane magnetic field \(H_z\) of the Z-shaped \(S=-2\) IS-PhCWG at \(a/\lambda = 0.283\), with transmittance 0.93. }
\label{fig:figure3}
\end{figure}
\subsubsection*{(ii) \(S=-2\), glide-symmetric waveguides}

Here we investigate the \(S=-2\) IA-PhCWGs without inversion symmetry (inset of Fig.3(a)). 
\textcolor{black}{As described before, this lattice and its domain-wall structure possess typical VPhC properties, such as non-trivial Berry curvature and angular momentum. In addition,}
due to the glide symmetry of this waveguide, two bands degenerate at the edge of the Brillouin zone (BZ), as shown in Fig.3(a). The modes have a mixed spatial parity at the waveguide's interface. The band above the degenerate frequency (upper band) has a broad single-mode region (\(a/\lambda = 0.267-0.301\)). However, the band below the degenerate frequency (lower band) overlaps with the bulk modes, making it impossible to excite the lower band. Here we focus on the upper band only. The upper band exhibits a very high transmission, as shown in Fig.3(b). \(T_{av}\) and \(R_{FP}\) in this Z-shaped waveguide are estimated to be 0.97 and 0.03. Figure3(c) shows the \(H_z\) distribution at \(a/\lambda = 0.270\), where the transmittance is 0.97. There is no indication of attenuation during the propagation. In addition, there is no apparent reflected intensity behind the excitation source, indicating very weak backscattering. These results show that the reflection at bends is significantly small. This configuration (S = -2) corresponds to a bearded interface in a honeycomb lattice VPhC waveguide \cite{Mref7_He2019_h16}. This result of high transmission is essentially similar to those reported in reference [19,23].

Next, we investigate \(S=-2\) waveguides with inversion symmetry by changing the air hole shape from triangular to circular. 
\textcolor{black}{That is, we keep the same lattice configuration but restore the inversion symmetry.} 
As shown in Fig.3(d), \(S=-2\) IS-PhCWG has a wide single-mode region (\(a/\lambda = 0.261-0.293\)) with a degeneracy point at the BZ edge (\(a/\lambda = 0.270\)). Since the degeneracy point is located within the bandgap, both upper and lower bands have sufficient single-mode regions. Surprisingly, the Z-shaped waveguide shows also very high transmission (Fig.3(b)) with \(T_{av}\) of 0.94 in the upper band, even though the inversion symmetry is NOT broken. The \(R_{FP}\) is 0.06. This high transmittance is comparable to that of the upper band in \(S=-2\) IA-PhCWG and other reported results in Z-shaped VPhCWGs\cite{Mref1_Majiawen2019_h23,Mref2_Yamaguchi_2019_h23,Mref3_Mikhail2019_h23,Mref4_ChenXD2017_h23,Mref4.1_HeXT2022_t13,Mref5_HAN2021_t16,Mref7_He2019_h16,Mref8_JalaliMehrabad2020_h56,Mref9_Yoshimi:21_h56,Mref10_Yoshimi:20_h56,Mref11_Cehn2019_h16,Mref12_Gao2017_h56,Mref13_Wu2017_s0,Mref14_Zhang2020_s0,Mref15_Kang2018_s0,Mref16_Ma2016_s0,Mref17_Yang2018_h23,Mref19_Zeng2020_s0,Mref20_DU2020_s0,Mref22_Dongjianwen2017_HCbulk,Mref23_XiangXi_HCzigzag,Mref24_Wang_2019_S0}. Figure3(c) shows the \(H_z\) distribution at \(a/\lambda = 0.283\) (upper band), where the transmittance is 0.93. Same as that of \(S=-2\) IA-PhCWG, there is no indication of attenuation during the propagation. The present result implies an important consequence. Because the inversion symmetry is not broken in this waveguide (S = -2 IS-PhCWG), it suggests that the observed high transmission may not be caused by the valley-photonic effect, which essentially requires the broken inversion symmetry.  

Interestingly, the transmittance of the lower band differs significantly from that of the upper band in \(S=-2\) IS-PhCWGs. The \(T_{av}\) of only 0.40. The \(R_{FP}\) is 0.52 which is even larger than that of the W1WG. It should be noted that similar transmission contrast between upper and lower modes has been reported for glide-symmetric honeycomb-lattice VPhC waveguides.  Yoshimi et al.\cite{Mref9_Yoshimi:21_h56} have reported a distinctive transmission contrast in a bearded-interface glide-symmetric PhCWG with a honeycomb lattice (S = 2 waveguide in the terminology of the present paper). Although the high bend-transmission was recently suggested for the upper band of an \(S=-2\) triangular-lattice IS-PhCWG\cite{Mref6_YangJK2021_c16}, it has not been reported that the similar high contrast between upper and lower modes exists in the triangular-lattice IS-PhCWG, which we believe is important for understanding the nature of the transmission through sharp bends.  Since the upper/lower band transmission contrast has been observed in Z-shaped 
\textcolor{black}{glide-symmetric PhCWGs irrespective of the inversion symmetry},
we speculate that this phenomenon originates from the domain-wall configuration instead of the symmetry-breaking in the bulk lattice.

\subsubsection*{(iii) Other \(S\)-value waveguides and summary of numerical studies}
\begin{table}[hbt!]
\includegraphics[width=1.0\textwidth]{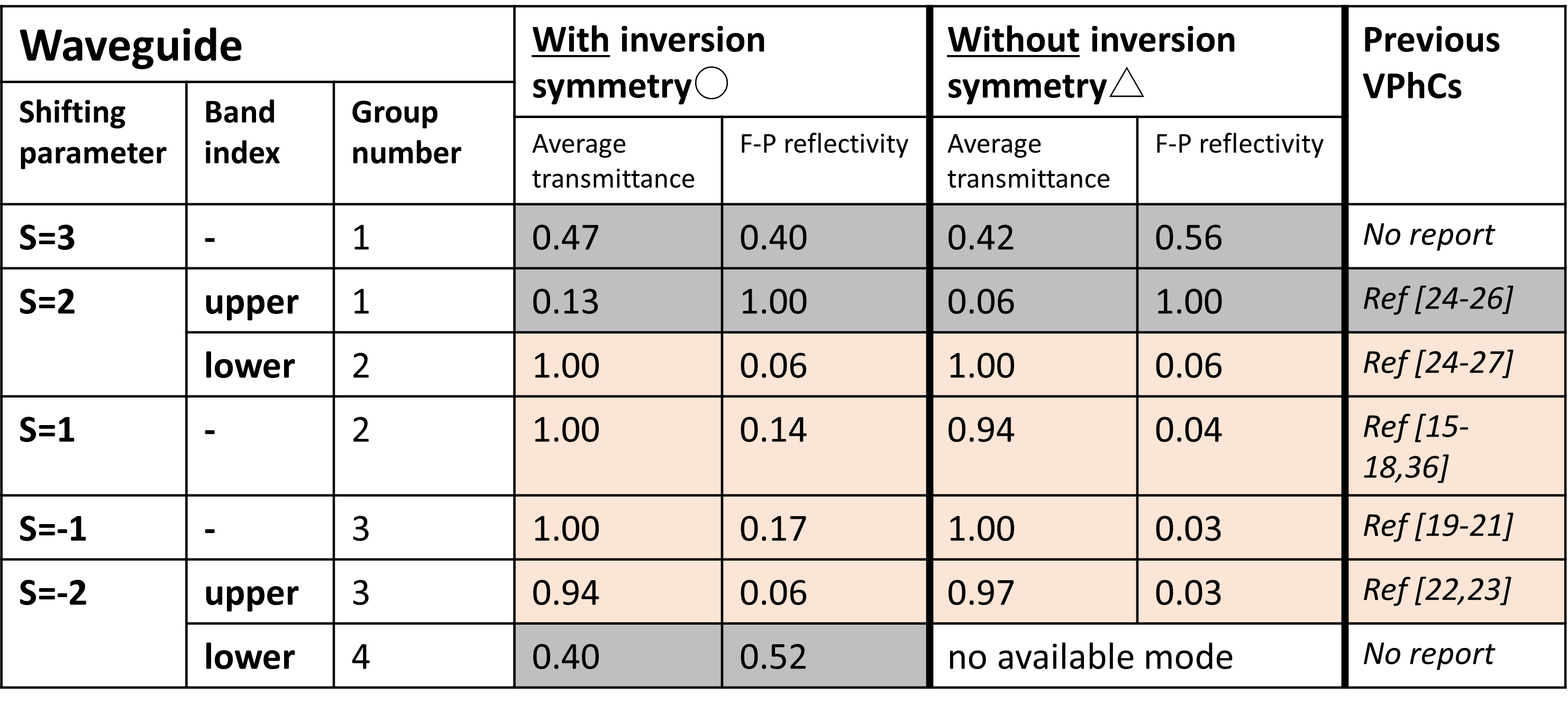}
\caption{The simulation results of average transmittances of the waveguide in Z-shaped waveguides. The rightmost column shows the previous reports. Orange text background indicates high bend-transmittance results, and gray text background indicates low bend-transmittance results. The group numbers will be explained in the discussion section.}
\end{table}
Following the same method as W1WGs and \(S=-2\) PhCWGs, we have also numerically investigated \(S=-1\) IA-PhCWGs, \(S=-1\) IS-PhCWGs, \(S=1\) IA-PhCWGs, \(S=1\) IS-PhCWGs, \(S=2\) IA-PhCWGs, \(S=2\) IS-PhCWGs, and \(S=3\) IA-PhCWGs. For these waveguide types, we only report a brief result here and leave a detailed discussion in the supplementary information (2).

 Table 1 summarizes \(T_{av}\) and \(R_{FP}\) for each \(S\)  with and without inversion symmetry. Bands with high bend-transmittance are labeled in orange and bands with low bend-transmittance are in gray. When \(S\) is even, the results of both the upper and the lower bands are shown. 

This table shows that \(S=1\), the upper band of \(S=-2\), and the lower band of \(S=2\) have high \(T_{av}\) and low \(R_{FP}\).  Most importantly, these characteristics do not depend on the existence of the inversion symmetry. It is worth noting that a large transmission contrast between the upper and lower bands for glide-symmetric waveguides, which was previously noted as proof of "topological property" for one of the bands, is also seen for inversion-symmetric waveguides. The last column in Table 1 shows the classification of the reported results for VPhC by the \(S\) parameter. All previous results of VPhCWG studies can be classified in the same table, and coincide with our result. This table strongly suggests that this high transmission behavior does not originate from the broken inversion symmetry, but possibly originated from the domain configuration. Before concluding this speculation, we check other possible causes. In Supplementary Table 1, we examine the parity of modes, the group refractive index, and the sign of the group velocity for each case. As shown in the table, there is no correlation between these variables and the observed distinctive difference in the bend-transmittance. Thus these variables cannot explain the present phenomenon. Consequently, our results strongly suggest that the observed high bent transmittance and low reflectivity are attributed to the domain-wall configuration.

\subsection*{Experimental studies of Z-shaped waveguides}

In this part, we experimentally examine the transmission properties of bent PhCWGs. We implement the PhC structures in Si slabs fabricated by highly accurate lithography and etching process. The fabrication details are described in Methods. We couple the waveguides to the incident light from a wavelength-tunable laser with 5dBm power and measure the transmission spectra from 1355 to 1640 nm.

Figure4(a) shows the optical microscope image of the fabricated Z-shaped waveguide. The TE-polarized light is guided through a silicon taper and is coupled to the PhCWG via a silicon nanowire. The three segments of Z-shaped waveguides have lengths of \(100a\), \(30a\), and \(100a\) respectively. Figure.4(b) shows the bending part of a \(S=-2\) IS-PhCWG and Fig.4(c) shows the straight part of a \(S=1\) IA-PhCWG. The length of the straight waveguides is \(230a\). In a straight waveguide, there are reflections between the PhCWG and the silicon waveguide, making the whole waveguide an F-P cavity. F-P resonances may occur inside both the \(100a\) segment and the \(30a\) segment in a Z-shaped waveguide. As a typical example in our measurement, for a waveguide with a=400 nm, and the waveguide mode with a group velocity of  \(0.1c\) and wavelength of 1500 nm, the wavelength FSR is 1.2 nm, 2.8 nm, and 9.4 nm in the \(230a\), \(100a\) and \(30a\) cavity respectively.
\begin{figure}[hbt!]
\includegraphics[width=0.9\textwidth]{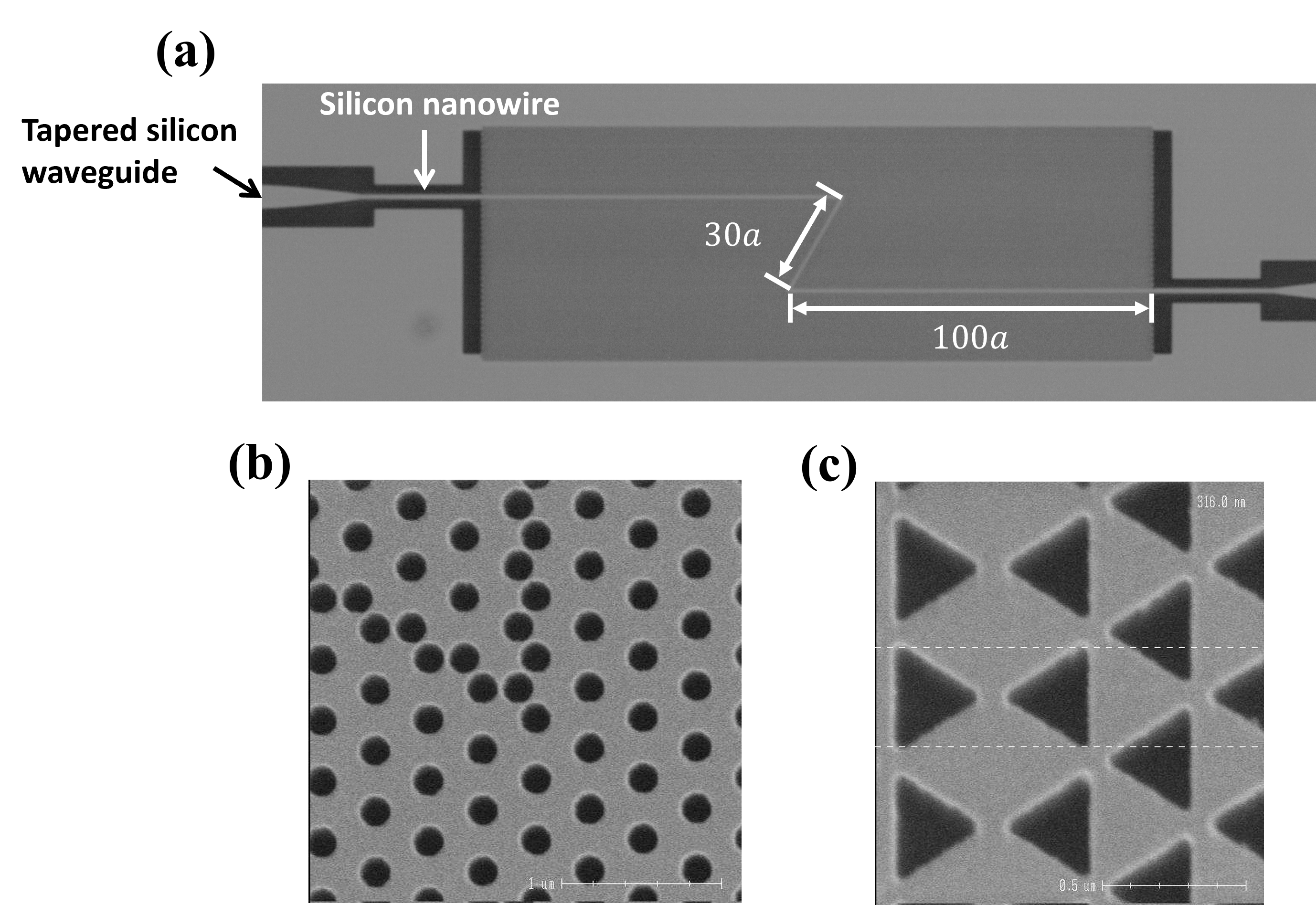}
\caption{(a) Optical microscope image of the fabricated Z-shaped \(S=3\) IS-PhCWG. The three segments of the waveguide have lengths \(100a\), \(30a\), and \(100a\). (c) SEM image of the bending part of a \(S=-2\) IS-PhCWG. (d) SEM image of the straight part of a \(S=1\) IA-PhCWG.}
\label{fig:figure8}
\end{figure}

Here we show the measured transmitted intensity. We begin with \(S=3\) IS-PhCWGs (W1WGs). The lattice constant \(a\) is 424 nm. The radius of air holes \(r\) is 97 nm. As shown in Fig.5(a), \(S=3\) IS-PhCWGs have single waveguide modes between 1515-1574 nm (yellow region). We evaluate \(T_{av}\) in the single mode region. For \(S=3\) IS-PhCWGs, \(T_{av}=0.31\). Note that this Z-shaped waveguide shows ripples in the spectrum, whose FSR is 3-10 nm. The observed FSR seems to roughly correspond to the FP resonance of the \(30a\) segment, but the ripple is complicated especially in the
longer wavelength region. Figure5(b) shows the spectra of \(S=3\) IA-PhCWGs (\(a\) = 416 nm), which have the same domain-wall configuration as W1WG but triangular air holes that break the inversion symmetry. The side length of the triangular air holes \(s\) is 232 nm. The \(S=3\) IA-PhCWGs have single modes between 1463-1532 nm. The evaluated \(T_{av}\) is also low, 0.48. These results agree with our theoretical simulation. Similar to \(S=3\) IS-PhCWGs, there are strong ripples in the spectrum. However, the ripples are rather complicated and hard to analyze. This trend is seen in all samples shown below. We regard that these complicated spectra result from the complex multiple reflection at section boundaries which exist in the fabricated devices but does not in the simulated structures. Hence, we could not evaluate \(R_{FP}\) from the measured spectra.
\begin{figure}[hbt!]
\includegraphics[width=0.9\textwidth]{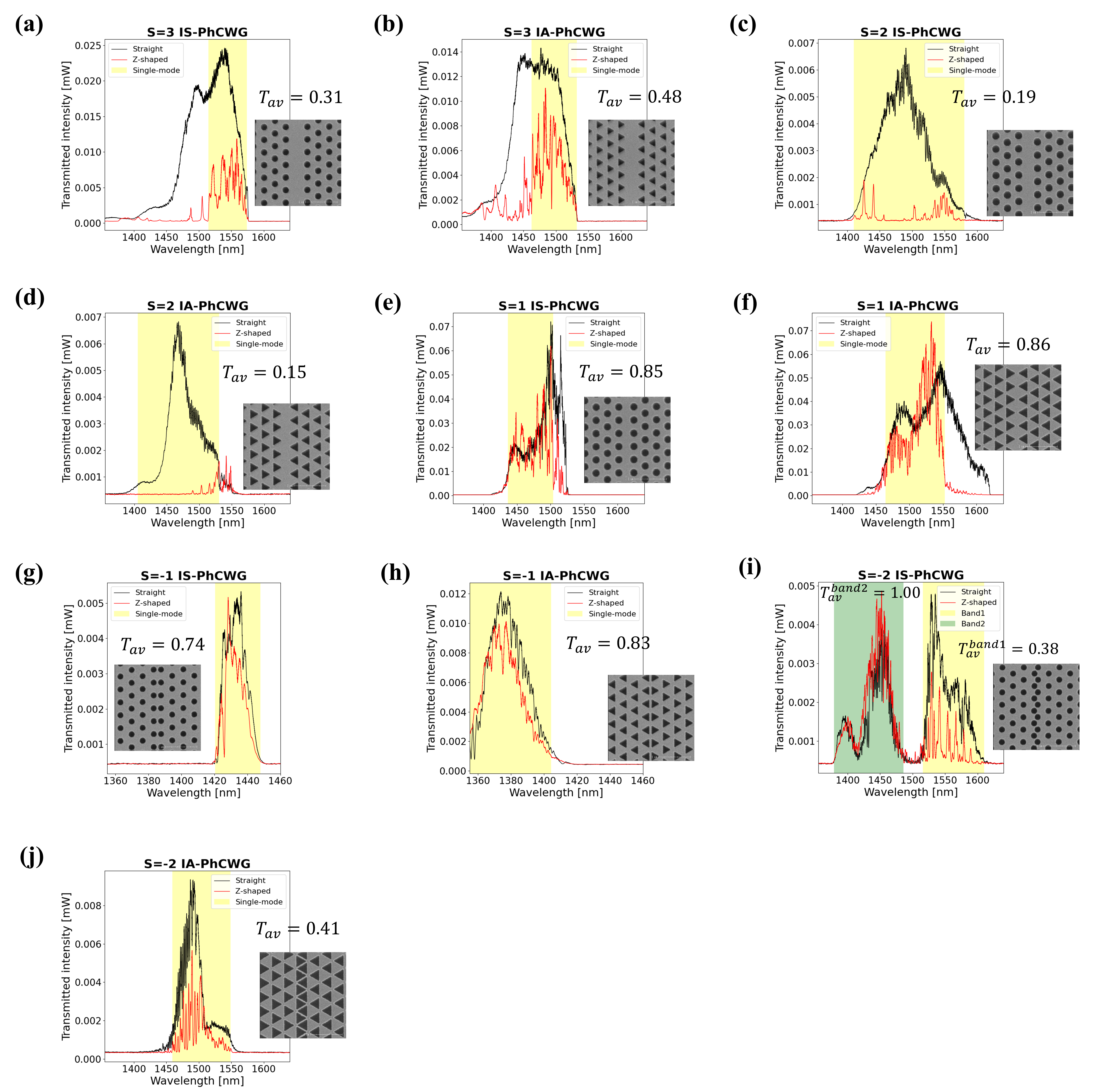}
\caption{Measured transmission spectra of (a,b) \(S=3\) PhCWG, (c,d) \(S=2\) PhCWG, (e,f) \(S=1\) PhCWG, (g,h) \(S=-1\) PhCWG and (i,j) \(S=-2\) PhCWG. Black lines indicate the straight waveguides and red lines indicate the Z-shaped waveguides with two bends. Yellow and green boxes show the single-mode regions. \(T_{ave}\) shows the relative average transmittance of the Z-shaped waveguides. A SEM image of the corresponding waveguide is shown on the right side of each plot.}
\label{fig:figure9}
\end{figure}

Hereafter, we investigate other domain-wall configurations one by one and evaluate \(T_{av}\). Figure.5(c) shows \(S=2\) IS-PhCWGs (\(a\) = 441 nm, \(r\) = 92 nm). As discussed in the simulation section, the \(S=2\) PhCWGs have glide-symmetric interfaces and have two touching bands in the photonic bandgap. However, the single-mode region (1410-1580 nm) only exists in the (frequency-wise) upper band. The upper band has an average bend-transmittance \(T_{av}\) of 0.19, lower than that of the W1WG.
Figure.5(d) shows the spectra of \(S=2\) IA-PhCWGs (\(a\) = 418 nm, \(s\) = 342 nm), which corresponds to a bearded interface in honeycomb lattice VPhCWG\cite{Mref12_Gao2017_h56,Mref9_Yoshimi:21_h56,Mref10_Yoshimi:20_h56,Mref8_JalaliMehrabad2020_h56, Mref25_Guillermo_disorderH23}. Like its inversion-symmetric counterpart, \(S=2\) IA-PhCWG has single modes (1405-1530 nm) in the upper band. The \(T_{av}\) is 0.15. The extremely low transmittance in the upper band of \(S=2\) PhCWGs agrees with the previous reports\cite{Mref12_Gao2017_h56,Mref9_Yoshimi:21_h56,Mref10_Yoshimi:20_h56,Mref8_JalaliMehrabad2020_h56} as well as our simulation results. 

Figure.5(e) shows the result for \(S=1\) IS-PhCWGs (\(a\) = 427 nm, \(r\) = 120 nm). The overall transmission in the Z-shaped \(S=1\) IS-PhCWGs (red) is comparable to that of the straight waveguide. \(T_{av}\) is as high as 0.85 in the single-mode range 1437-1504 nm.
\(S=1\) IA-PhCWGs (\(a\) = 441 nm, \(s\) = 314 nm) corresponds to a zigzag interface VPhCWG. The previous reports are all based on the honeycomb lattice \cite{Mref17_Yang2018_h23,Mref4_ChenXD2017_h23,Mref3_Mikhail2019_h23,Mref1_Majiawen2019_h23,Mref2_Yamaguchi_2019_h23,Mref18_Chen2021_sandwiched_h23}. Similar to its IS- counterpart, the Z-shaped \(S=1\) IA-PhCWGs have transmission comparable to that of the straight waveguides (Fig.5(f)). \(T_{av}\) is 0.86 in the single-mode region 1464-1553 nm.

Figure.5(g) shows the measured spectra of \(S=-1\) IS-PhCWG (\(a\) = 419 nm, \(r\) = 92 nm) has a narrow single-mode region from 1420 to 1448 nm. The \(T_{av}\) is 0.74. The amplitude of F-P resonances in the Z-shaped \(S=-1\) IS-PhCWG's spectrum is as small as that of the straight waveguide. \(S=-1\) IA-PhCWG (Fig.5(h), \(a\) = 442 nm, \(s\) = 268 nm)) corresponds to another type of zigzag VPhCW\cite{Mref4.1_HeXT2022_t13}. It also has a narrow single-mode region from 1355 to 1404 nm. The \(T_{av}\) is 0.83.

Finally, we investigate the \(S=-2\) PhCWGs with glide-symmetry interface. Figure.5(i) shows the result for \(S=-2\) IS-PhCWGs (\(a\) = 418 nm, \(r\) = 115 nm). Both the straight and the Z-shaped waveguides have a transmission gap near 1500 nm. We speculate that fabrication errors in the air holes' size break the glide-symmetry, opening a bandgap and creating flat-band regions near the edges of the upper/lower bands. The upper band (green, 1379-1486 nm), shows higher transmission in the Z-shaped waveguide than in the straight waveguide. Therefore we set the \(T_{av}\) to be 1.00. The lower band (yellow, 1516-1610 nm) shows low transmittance in the Z-shaped waveguide. The \(T_{av}\) is 0.38. 
Figure.5(j) shows the spectra of the \(S=-2\) IA-PhCWGs (\(a\) = 469 nm, \(s\) = 287 nm). In the numerical calculations, only the upper band of the \(S=-2\) PhCWGs have single modes. Actually, the lower band also lies within the bandgap in the 3-dimensional device. Here only the lower band can be observed possibly owning to the fabrication errors that separate the two bands with a large bandgap. Like the \(S=-2\) IA-PhCWGs, the lower band has low transmittance in the Z-shaped bends. The \(T_{av}\) is 0.41.

To summarize, the experimental results clearly show a significant difference in the transmission properties among different domain-wall configurations. The waveguide modes in \(S=1\), \(S=-1\) PhCWGs, and the upper band of \(S=-2\) PhCWG have high bend-transmittance, which agrees well with the numerical calculation results. Our experimental results support our proposal that the bend-transmittance is dominantly determined by the domain-wall configuration.

\section*{Discussion}
We have numerically and experimentally demonstrated high bend-transmission through 120-degree sharp bends for PhCWGs with various \(S\) regardless of the existence of the inversion symmetry. In addition, the high transmission appears regardless of the mode parity and the group velocity. To observe the correlation between \(S\) and the bend-transmittance more intuitively, we plot \(T_{av}\) against \(R_{FP}\) for each simulation result in Fig.5(a). If \(S\) changes continuously, the waveguide bands will accordingly evolve in the bandgap, until they disappear into the bulk mode regions. Different bands can be traced to one another in this process before disappearing. We classify such bands into the same group as \(S\) changes from 3 to -2 (see supplementary (5) for detailed explanations). Thus we can classify the investigated waveguide bands into four different groups. We label each band group with different colors in Fig.5. The even mode of W1WG and the upper band of \(S=2\) glide-symmetric PhCWG belong to group 1 (green). The lower band of \(S =2\) glide-symmetric PhCWG and the even band of \(S=1\) PhCWG belong to group 2 (blue). The odd mode of \(S=-1\) PhCWG and the upper band of \(S=-2\) glide-symmetric PhCWG belong to group 3 (orange). Finally, the lower band of \(S=-2\) glide-symmetric PhCWG belongs to group 4 (gray). 

Now we focus on the transmission properties of each band group. As shown in Fig.5(a), group 2 (blue) and group 3 (orange) have high bend-transmittance and low F-P reflectivity, and group 1 (green) and group 4 (gray) have low bend-transmittance and high F-P reflectivity. This shows that the mode classification corresponds with the transmission property of the Z-shaped waveguide very well. In addition, each group includes waveguide bands of both IS-PhCWGs and IA-PhCWGs, meaning that the existence of inversion symmetry has no significant influence.
\begin{figure}[hbt!]
\includegraphics[width=1.0\textwidth]{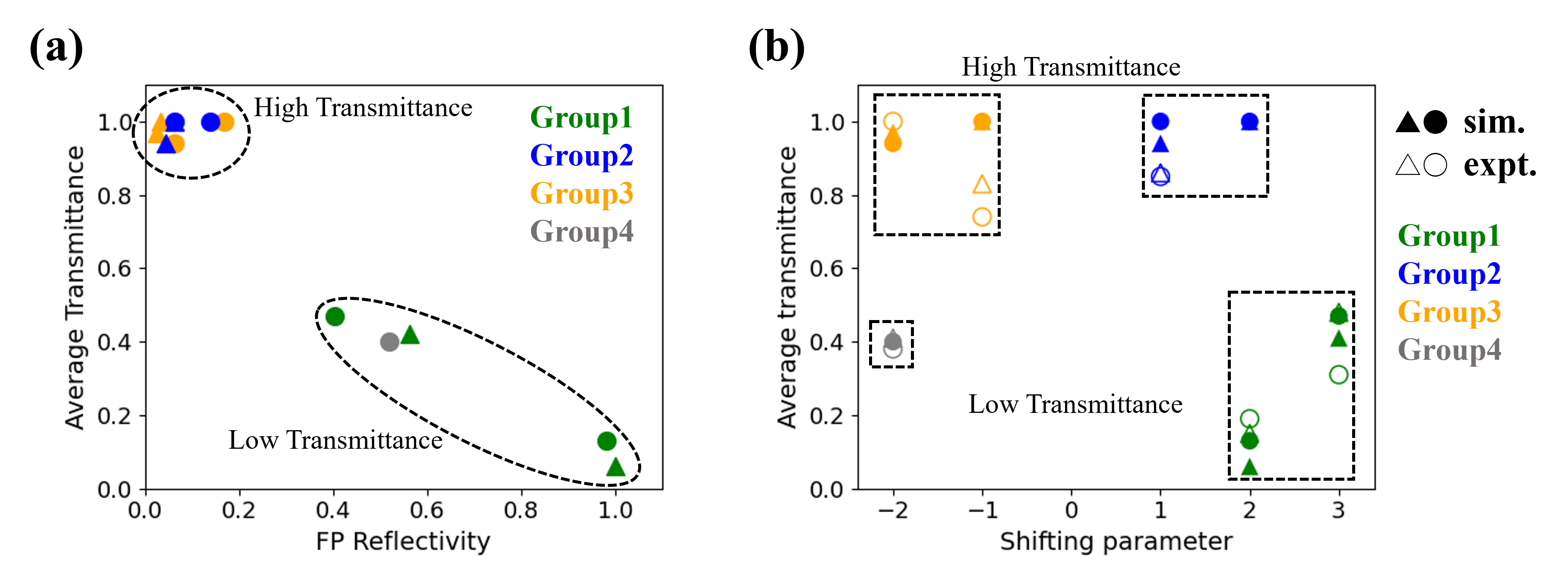}
\caption{(a) Average transmittance against F-P reflectivity for simulation results. The circles indicate the results of waveguide modes in IS-PhCWGs, and the triangles indicate that of IA-PhCWGs. Colors indicate the mode classification. Black texts show the shifting parameters of each waveguide, where U indicates the upper band, and L indicates the lower band for a glide-symmetric interface. (b) Average transmittance against shifting parameters for simulation and experiment results. Solid markers indicate the simulation results. Hollow markers indicate the experiment results.}
\label{fig:figure11}
\end{figure}
 It is worth noting that when continuously changing \(S\), groups 1 and 2 are connected at the BZ edge degeneracy point of the \(S=2\) PhCWG, and groups 3 and 4 are connected at the BZ-edge degeneracy point of the \(S=-2\) PhCWG. Interestingly, a degeneracy point of glide-symmetric waveguides connects two different band groups having high and low transmissions. 
 %In addition, groups 2 and 3 show similar transmission properties, which are not connected by the degeneracy point during a continuous change of \(S\), but separated by the bulk bandgap of \(S = 0\) (no lattice shifting).

 Next, we summarize the \(T_{av}\) obtained in simulation and measurement in Fig.6(b). We plot \(T_{av}\) against the shifting parameters. Solid markers show the simulation results and hollow markers show the experiment results. The experiment results agree well with the simulation results, showing that group 1 and group 4 have low bend-transmittance, and group 2 and group 3 have high bend-transmittance.

So far, we have clarified that the high bend-transmission is not influenced by the bulk property (especially for the existence of the inversion symmetry), and is mostly determined by the domain-wall configuration. This result has already denied the conventional understanding of the high bend-transmission as the valley-photonic property, in which the transmission is determined by the bulk topological property.  More importantly, our result shows that high bend-transmission can occur for a much wider range of PhCWGs, which is not restricted to inversion-asymmetric PhCs. This finding is promising considering applications. 

We further discuss the mechanism behind the mode classification. It has been clarified that one can find topologically-stable polarization singularities, such as the circular polarization singularities (C-points or CPs) and vortex singularities (V-points or VPs)\cite{CP-kuiper,CP-ABYoung,Sollner2015,CP-disorder} of the electric/magnetic fields. Unidirectional excitation of waveguide modes can be realized by putting the circularly polarized wave source at the location of the CPs\cite{Mref8_JalaliMehrabad2020_h56,Sollner2015,Mref6_YangJK2021_c16,CP-ABYoung}. CPs and VPs exist pervasively in many types of PhCWGs regardless of the band topology or the particular symmetry of the structure. Here we speculate that the transmittances via sharp bends are related to the spatial distribution of CPs and VPs in the PhCWGs. 

\begin{figure}[hbt!]
\includegraphics[width=1.0\textwidth]{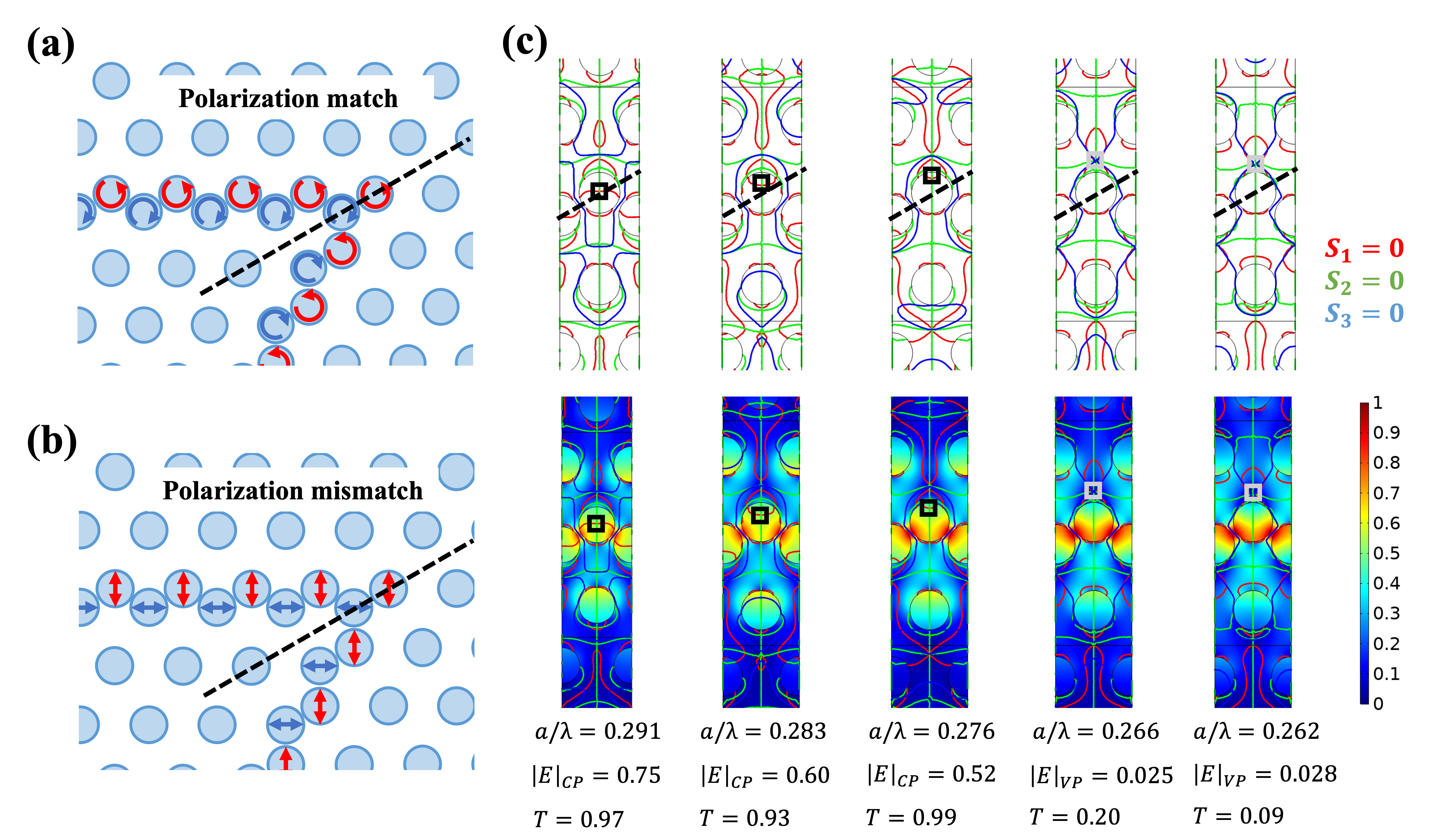}
\caption{Schematic illustration of (a) polarization singularities causing high transmittance and  (b) linear polarization causing low transmittance in the bend of an \(S=-2\) IS-PhCWG. The red and blue arrows represent the directions of polarization. Black dotted lines indicate the connection interface of the 120-degree bend. (c) The simulation results of the location of polarization singularities. Red, green, and blue lines show the zero-values of \(S_1\),\(S_2\), and \(S_3\), respectively. Black boxes show the location of CPs. Gray boxes show the location of VPs. Color maps show the amplitude of electric fields in the waveguide. Black dotted lines show the connection interfaces of a 120-degree bend. The normalized frequencies, the normalized amplitudes of the electric field, and the bend-transmittances are shown below each plot.}
\label{fig:figure11}
\end{figure}

When the light is circularly polarized at the mirror-symmetric line of a 120-degree bend (broken lines in Fig.7(a)), the polarization state would be preserved across the bending, i.e., changing the propagation direction by 120 degrees does not cause polarization mismatch. Otherwise, If the light is not linearly polarized along this line, the polarization state is not preserved after the bending, and polarization mismatch occurs when constructing bends, which leads to a large modification of the field profile and may generate unwanted local resonances and/or high reflection (Fig.7(b)). Thus, circular polarization may lead to a smooth connection and thus small reflection within a broad frequency range.  Here we examine this speculation by analyzing the polarization profile of the simulation results and investigating the distribution of CPs. In Fig.7(c), we plot the CP distributions in unit cells of \(S=-2\) IS-PhCWG at several frequencies. To identify the location of CPs, we plot the zero-value isolines of Stokes parameters\cite{CP-disorder,CP-poincare}. The crossing nodes of the \(S_1=0\) (red) and\(S_2=0\) (green) lines indicate C-points (CPs). The crossing nodes of the \(S_1=0\), \(S_2=0\), and \(S_3=0\) (blue) lines indicate VPs.  

The upper band of \(S=-2\) IS-PhCWG has high bend-transmittance, and there are CPs (black boxes) near the broken lines. The calculated degree of polarization (\(S_3/S_0\)) is over 0.99 at each CP. The normalized electric field amplitudes (\(I/I_{max}\)) are over 0.5 at each CP, i.e., these are bright CPs. These CPs are located inside the air holes. As we change the frequency, we observe that these CPs gradually move away from the hole centers and disappear exactly at the degeneracy point. 
Beyond the degeneracy point, the lower band appears, which has low bend-transmittance. In this regime, as shown in the fourth and fifth columns in Fig.7(c), there are no bright CPs near the broken lines. There are only dark CPs in the silicon area where \(I/I_{max}\) is lower than 0.1. In addition, VPs (gray boxes) appear near the broken lines. The sudden disappearance of bright air-hole CPs is consistent with the abrupt change in bend-transmittance across the degeneracy point. 

We have also investigated other domain-wall types of IS- and IA- PhCWGs (supplementary information (6)). We have confirmed the existence of bright air-hole CPs in most of the high bend-transmission bands and their disappearance in the low bend-transmission bands. For the glide-symmetric \(S=2\) PhCWGs, we have also observed that CPs located inside the air holes disappear around the position where the bend-transmission abruptly decreases. These results suggest that there is some correlation between the high bend-transmission and the existence of CPs near the mirror-symmetric line. Another important point is that for all the investigated domain walls, the CP distribution in IA- and IS- PhCWGs have no significant difference. This means that the inversion symmetry does not alter the CP properties. We admit that these arguments are still speculative because it is difficult to estimate the transmittance quantitatively from the distribution and brightness of CPs. We leave detailed investigations in this direction for future works.

As a final remark, we would like to address the influence of breaking inversion symmetry on bend-transmissions. Notably, we have observed a marginal enhancement in bend-transmission for certain IA-PhCWGs when compared to IS-PhCWGs, as evidenced by both numerical calculations and experimental data. It is important to note that this modest variation should be distinguished from the primary transmission contrast discussed in this study. Nevertheless, there exists the possibility that this slight improvement can be attributed to the valley-photonic effect. However, it is currently beyond the scope of this study to conduct a quantitative analysis of this subtle difference with our waveguide design, and we leave it for future works.
Since the recent works \cite{Mref25_Guillermo_disorderH23,disroder2_Rosiek2023} indicate that the suppression of the backscattering may occur in a slow light region with a small disorder, one needs to analyze this issue in sharp bends in a meticulous manner.

\section*{Conclusion}
In summary, we have investigated the bend-transmission in a series of triangular-lattice PhCWGs compatible with 120-degree sharp bends. We systematically investigated different domain-wall configurations by adjusting \(S\) for waveguides with and without the inversion symmetry. Our numerical and experimental results demonstrate that significantly high bend-transmission can be achieved for certain domain-wall types, including typical VPhCWGs. Surprisingly, the presence of the inversion symmetry does not affect the emergence of high bend-transmission, which contradicts the previous understanding of the VPhCWGs. Our findings provide new possibilities for achieving uniquely high bend-transmission in a broader range of PhCs, not restricted to VPhCWGs. Since bending loss is one of the serious issues for nanophotonic integrated circuits, this work carries significant implications for constructing flexible low-loss nanophotonic circuits. As an empirical explanation, we propose a mode classification that links the high bend-transmission to specific groups of waveguide modes. Regarding the origin of the high bend-transmission, a preliminary study suggests that the abrupt change in bend-transmission is accompanied by the emergence or disappearance of topologically-protected CPs near the bending interface. Therefore, we speculate that the high bend-transmission phenomenon is related to the existence of CPs at the interface, whose behavior is mostly determined by the domain lattice configuration and is minimally influenced by the presence of the inversion symmetry. It remains possible that the observed slight difference in the bend-transmission between the PhCWGs with and without the inversion symmetry can be attributed to the suppression of backscattering due to the valley-photonic effect. However, an unambiguous conclusion requires more detailed and deliberated work. Our present work may pave the way to open up novel designs of nano-waveguides for low-loss nanophotonic integrated circuits and shed new light on the nature of valley-photonic properties.    

\section*{Method}
\subsection*{Simulations}
We implement our waveguide designs on the SOI platform, using the transverse-electric (TE) modes confined in the PhC slab. The thickness of the slab is 220 nm. The lattice constant is 400 nm. The radii of circular air holes are 102 nm, and the length of one side of the triangular air holes is 277 nm. The area of each circular and triangular air hole is approximately the same. We conduct simulations based on a finite element method using commercial software (COMSOL). We first calculate the three-dimensional (3D) photonic band structure (PBS) and then approximate the 3D PBS in two-dimensional (2D) models. In 3D calculation, the refractive index of silicon is set as 3.48. In 2D calculation, the effective refractive index of silicon is set to be 2.65 to keep the photonic bandgap (PBG) within approximately the same wavelength range as the 3D results. As shown in Fig.1, we confirmed that both IS- and IA-PhCs have a broad PBG of over 250nm. 

By connecting the bulk PhCs with an arbitrary interface, we can construct PhCWGs that support interface modes. The broken inversion symmetry that causes the coupling between the chirality of modes and valley DoF is the foundation of the valley-photonic explanation of high transmission in sharp bends. This explanation will remain valid if we observe low transmission in IS-PhCWGs and high transmission in IA-PhCWGs. Otherwise, if we observe relatively high transmission in some IS-PhCWGs or relatively low transmission in some IA-PhCWGs, we should consider other factors affecting the transmission other than inversion symmetry.
For each domain-wall type in Fig.1(c), we construct the IA-PhCWGs and the IS-PhCWGs. We connect domains constructed from patterns A and B to construct the IA-PhCWGs with broken inversion symmetry. In supplementary information (1), we have numerically confirmed that the A-B and B-A type waveguides have no essential difference in their transmission properties. Here we set most of the IA-PhCWGs to be the A-B type interface for convenience. \(S=-1\) IA-PhCWGs are exceptions and have a B-A type interface because the air holes overlap with each other at the A-B type interface. We calculate the light transmittance through a straight waveguide and a Z-shaped waveguide of the same length for each type of waveguide design. Details of the settings of the wave source are described in the supplementary information (1).

For each waveguide band, we calculate the average transmittance and estimate the reflectivity at bends from the amplitude of the F-P ripples.
Given the bend reflectivity R, the transmittance through an F-P cavity is \(T=\frac{1}{1+F(\sin\frac{\sigma}{2})^2}\), where \(F=\frac{4R}{(1-R)^2}\) is the finesse factor. Therefore, we can derive the reflectivity R for the transmittance spectra.
We call the calculated R the F-P reflectivity and use this value to evaluate the bend-transmission in addition to the average transmittance. 

Undesirable mode conversions may occur if other waveguide modes or bulk modes are near the edge of the single-mode region. In addition, we disregard the ultraslow light region near the mode edge where large reflection makes the analysis difficult. Thus, the actual frequency range where the transmittance can be accurately evaluated is slightly narrower than that calculated in the PBS.
In our calculation, the frequency range in which the average transmittance and F-P reflectivity are calculated is set to be narrower than the single mode region of the waveguide band by 6 THz (\(0.08a/\lambda\)) when the lattice constant is 400 nm).

\subsection*{Fabrication and experiments}
We implement our waveguide design in 220nm-thick silicon slabs on 3000nm-thick air-bridge-structured $\text{SiO}_{\text{2}}$ under-claddings. The PhCWG patterns are fabricated by electron beam lithography and dry etching technique. The air bridge is formed by removing the sacrificial layer of $\text{SiO}_{\text{2}}$ with hydrofluoric acid. 

To measure the transmittance spectrum, light from a wavelength-tunable laser is launched into an input silicon waveguide with a width of \(8 \mu m\). The output intensity is 5 dBm. The silicon waveguide is connected with an appropriate taper to a silicon nanowire, which is 400-700 nm in width. The silicon nanowire is straight in the straight PhCWG and is \(12.5 \mu m\) in length. Light is coupled to the PhC region via the silicon nanowire. The transmitted light is collected from an output silicon waveguide of the same design. Due to fabrication errors, the transmitting wavelength ranges in most waveguides deviate from those predicted by the 3-dimensional band calculation. Therefore we determine the single-mode region directly from the transmission spectra. Considering that multi-modes can have relatively high transmission through a straight waveguide, we determine the cut-off wavelength of the single modes using the Z-shaped waveguide's spectra. Suppose the peak transmitted intensity is \(I_{max}\) at wavelength \(\lambda_{max}\) we defined the single-mode range as (\(\lambda_{1},\lambda_{2}\)) where \(\lambda_{1}=max(\{\lambda|\lambda<\lambda_{max},\:I(\lambda)<0.1I_{max}\})\), and \(\lambda_{2}=min(\{\lambda|\lambda>\lambda_{max},\:I(\lambda)<0.1I_{max}\})\).For waveguide bands that have extremely low transmittance via 120-degree bends, like the upper band of S=2 PhCWGs, we calculate (\(\lambda_{1},\lambda_{2}\)) using the straight waveguides' spectra in the same manner.

To eliminate the influences of insertion loss and coupling loss, we use the measured intensities of straight PhCWGs as the reference to calculate the average transmittance of the bent PhCWGs. We convert the transmitted intensities of the straight and bent waveguides into the linear scale (in milliwatts) and respectively calculate their average intensities in the single-mode region. The average transmittance of the bent waveguide is derived as the division of the bent waveguides' average intensity and the straight waveguides' average intensity. We may also calculate the relative transmittance of the bent waveguides before taking the average. However, due to fluctuations in the spectra, the bent waveguides' intensities can be larger than that of the straight waveguides' at some wavelengths, which is amplified in linear scale and brings unnecessary errors to the results. Most of the measured transmission spectra of the Z-shaped waveguides have complicated resonance ripples in addition to the simple pattern of the  \(30a\) cavity F-P resonance. Therefore it is difficult to calculate the FP reflectivity from the measured data in the same manner as in the numerical studies.
\newline

\noindent
\textbf{Data availability.} The data which support the figures and other findings within
this paper are available from the corresponding authors upon request.

\section*{Acknowledgement}
The authors would like to thank Masato Takiguchi for his help in the experimental setup, and Toshiaki Tamamura for his help in the fabrication process.
This work was supported by the Japan Society for the Promotion of Science (Grant number JP20H05641) and the Japan Science and Technology Agency (JST Spring, Grant Number JPMJSP2106). 

\section*{Author contributions}

M.N. conceived the ideas. M.N. and Y.M. supervised the project.  M.N. and T.Y. proposed the theoretical background. W.D., M.O., and E.K. conducted the fabrication process. W.D. conducted the simulations and experimental measurements. Y.M. and T.Y. helped with simulations. W.D. and M.N. wrote the manuscript with feedback from other authors. 

\section*{Additional information}

\textbf{Supplementary Information}: Supplementary Information accompanies this paper at doi:\newline

\noindent
\textbf{Competing interests}: The authors declare no competing financial interests.

\bibliography{main-ref}

\end{document}

\flushbottom
\maketitle

\section*{1. Simulation method}
As shown in Fig.1(a) the PhCWGs are surrounded by perfectly matched layers (PMLs). The total length of each waveguide is 100a. A unidirectional dipole source is put inside each waveguide, perpendicular to the waveguide direction, to excite the waveguide modes. The width of the dipole source is set to 2a. We have confirmed that changing wave source width does not affect the simulation results. The source is a combination of surface electric current and surface magnetic current. 
Surface current density \(Js =(0,\sqrt{2n_{eff}}/\sqrt{Z_0L_{dipole} },0)\)
Surface magnetic current density: \(Jms =(0,0,\sqrt{2Z_0}/\sqrt{n_{eff}L_{dipole} })\)
where \(n_{eff}\) is the effective refractive index, \(L\) is the dipole length and \(Z_0\) is the impedance in free space. 
By fixing the ratio \(Jms/Js=Z_0/n_{eff}\), we excite an EM wave propagating only rightwards.

\begin{figure}[tbh!]
\includegraphics[width=0.7\textwidth]{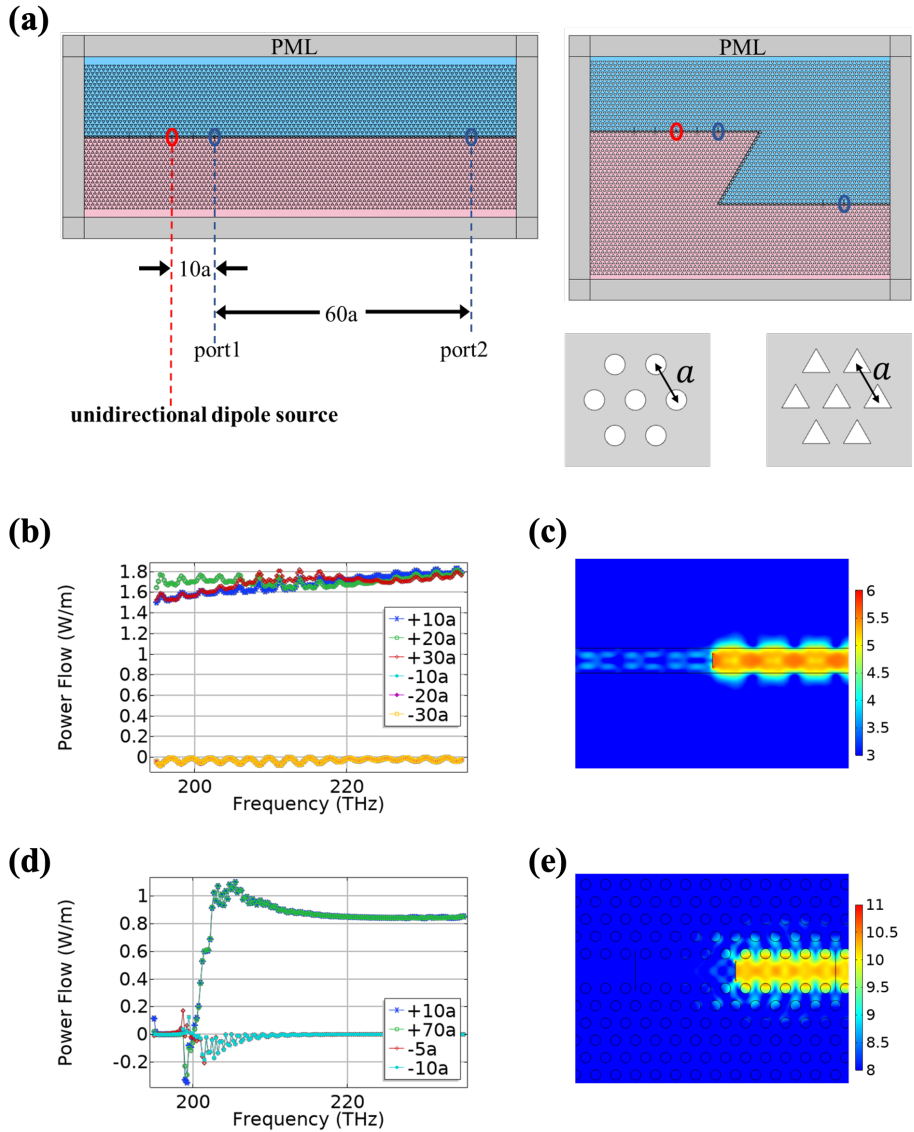}
\caption{(a) Schematic illustration of straight band bend waveguides in numerical calculation. The unidirectional dipole sources are positioned inside the red circles. 2 observation ports, port-1, and port-2 are positioned inside blue circles. The PhCWGs are surrounded by PMLs. Left is the straight waveguide and right is the Z-shaped bent waveguide with two 120-degree sharp bends. The bottom right is the enlarged bulk triangular lattice with and without inversion. (b) Power flow spectra of a silicon waveguide. The excited wave is assumed to propagate right-wards. 3 ports are set at each of the left and right sides of the dipole wave source, each 10a, 20a, and 30a away from the source. (c) The E-field intensity of the excited wave in the silicon waveguide at 200THz. (d) Power flow spectra of the W1WG are discussed in the simulation results. (e) The E-field intensity of the excited wave in the W1WG at 210THz}
\end{figure}
\clearpage

In the straight waveguide, 2 observation ports, port-1, and port-2 are located 10a and 70a away from the wave source. In the bent waveguide, port-1 is located 10a away and the first sharp bend is 30a away from the wave source. The second sharp bend is 20a away from the first one. And port-2 is 20a away from the second sharp bend. Thus, the observed EM waves travel the same distance between port-1 and port-2 in straight and bent waveguides. The source power \(P_0\) is calculated as the total power flow (Poynting vector) orthogonal to port-1 in the straight waveguide. The transmitted power in a straight waveguide \(P_s\) and that in the Z-shaped waveguide \(P_z\) are calculated as the total power flow orthogonal to port-2 at each waveguide. The transmittance is calculated as \(P_s/P_0\) in straight waveguides and \(P_z/P_0\) in Z-shaped waveguides. We cannot calculate the bend-transmittance naively as \(P_z/P_1\) where \(P_1\) is the total power flow passing through port-1 in the Z-shaped waveguide. The reason is that \(P_1\) also collects the power of reflected waves at the bends and is strongly affected by the F-P resonance. 
To examine the unidirectionality of the dipole source, we first confirm the wave propagation in a rectangular silicon waveguide without photonic crystal. The waveguide width is \(\sqrt{3}a\). As shown in Fig.1(b), the energy flow observed at the left-side ports (-10a, -20a, and -30a) is less than \(10\%\) of that observed at the right-side ports (10a,20a,30a). The E-field intensity at 200THz is plotted in Fig.1(c) in logarithm scale. We can see the majority of the energy propagates right-wards. In Fig.1(d,e), we have confirmed the same unidirectionality of a straight W1WG.

%In most waveguides we have investigated, the backward energy flow is sufficiently small to be neglected. Exceptions are the \(S=-1\) IS-PhCWG, \(S=1\) IS-PhCWG, where the excited EM waves do not couple efficiently with the waveguide mode. Therefore, an unneglectable portion of excited energy is lost and not received by the transmission ports. As a result, the calculated bend-transmittance has a maximum value larger than 1 and the averaged transmittance also exceeds 1. The F-P resonances, on the other hand, are unaffected by bad coupling and can still be confirmed from the spectra. The F-P ripples have steady amplitudes and the free spectra range corresponds to the cavity length. The Hz distribution also displays no large scattering at the bends. Therefore, despite this inaccuracy caused by bad coupling efficiency, we can still conclude that these two types of waveguides have good bend-transmission. 

\section*{2. Other simulation results}
We have reported in Table.1 and Fig.6(a) in the main manuscript the simulation results of \(S=-1\), \(S=1\), \(S=2\) IS- and IA- PhCWGs, and \(S=3\) IA-PhCWGs. \(S=-1\) and \(S=1\) PhCWGs have high transmissions, and \(S=2\) and \(S=3\) PhCWGs have lob transmissions via 120-degree sharp bends. Here we discuss these results in detail.
\begin{figure}[hbt!]
\includegraphics[width=1.0\textwidth]{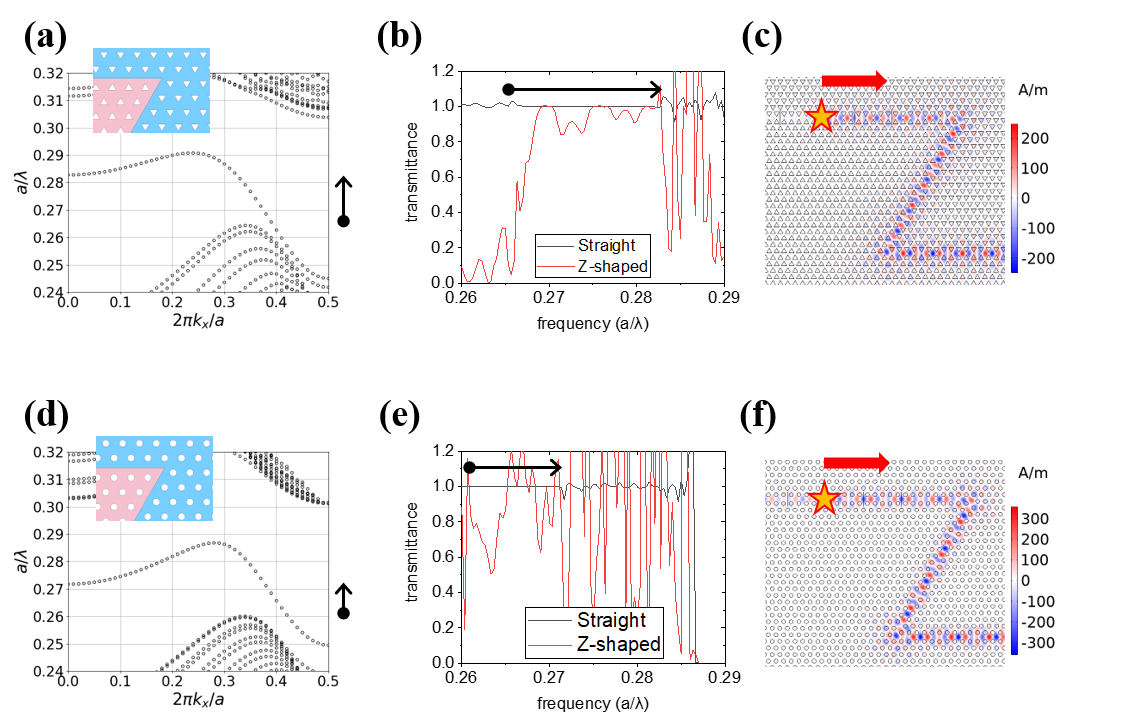}
\caption{Calculation results of the straight and Z-shaped \(S=1\) PhCWGs. (a) PBS of the \(S=1\) IA-PhCWG. (b) The transmittance spectra of the straight \(S=1\) IA-PhCWG (black curve) and Z-shaped bent \(S=1\) IA-PhCWG (red curve). The \(T_{av}\) is 0.94. The \(R_{FP}\) is 0.04. (c) The out-of-plane magnetic field \(H_z\) of a Z-shaped \(S=1\) IA-PhCWG at \(a/\lambda = 0.272\) with a transmittance of 0.88. (d) PBS of the \(S=1\) IS-PhCWG. The black arrow shows the single-mode region. The inset shows one corner of the bent waveguide. (e) The transmittance spectra of the straight \(S=1\) IS-PhCWG (black curve) and Z-shaped bent \(S=1\) IS-PhCWG (red curve). The \(T_{av}\) is 1. The \(R_{FP}\) is 0.14. (f) The out-of-plane magnetic field \(H_z\) of a Z-shaped \(S=1\) IS-PhCWG at \(a/\lambda = 0.267\) with transmittance of 0.90. }
\label{fig:figure4}
\end{figure}
\subsection*{(i) \(S=1\), mirror-symmetric waveguides}
 \(S=1\) waveguides have zigzag interfaces. This domain-wall configuration appears in many previous VPhC studies, reporting the high transmission in Z-shaped \(S=1\) PhCWGs with honeycomb lattice\cite{Mref1_Majiawen2019_h23,Mref2_Yamaguchi_2019_h23,Mref3_Mikhail2019_h23,Mref4_ChenXD2017_h23}. As shown in Fig.2
(a,d), the single-mode region is 0.260-0.272 in the IS-PhCWG and 0.266-0.283 in the IA-PhCWG. As shown in Fig.2(b,e), the transmittances of Z-shaped \(S=1\) IS-PhCWG and \(S=1\) IA-PhCWG are both very high compared to that of the W1WG. The \(S=1\) IS-PhCWG has \(T_{av}\) of 1.00 and \(R_{FP}\) of 0.14. The \(S=1\) IA-PhCWG has \(T_{av}\) of 0.94 and \(R_{FP}\) of 0.04. Note that the bend-transmittance of \(S=1\) IS-PhCWG is greater than 1 at some frequencies and averages 1.00 despite strong fluctuation in the spectrum. This is due to bad coupling between the waveguide mode and the wave source and the subsequent complicated F-P resonance. We can see from Fig.2(c) that the waveguide mode is well confined in the \(S=1\) IS-PhCWG, and the majority of wave flux travels through the Z-shaped bends without much scattering. Despite the technical issues in the numerical calculation, we can conclude that the \(S=1\) IS-PhCWG has good bend-transmission comparable to that of the \(S=1\) IA-PhCWG and \(S=3\) PhCWGs.

\begin{figure}[hbt!]
\includegraphics[width=1.0\textwidth]{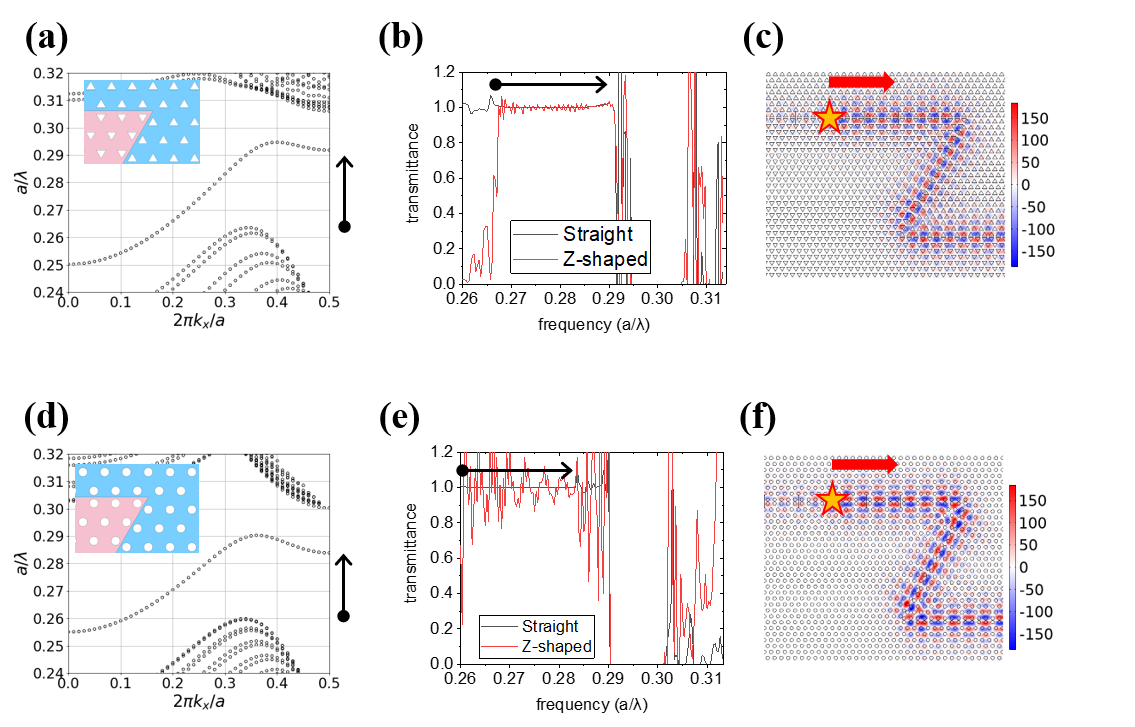}
\caption{Calculation results of the straight and Z-shaped \(S=-1\) PhCWGs. (a) PBS of the \(S=-1\) IA-PhCWG. (b) The transmittance spectra of the straight \(S=-1\) IA-PhCWG (black curve) and Z-shaped bent \(S=-1\) IA-PhCWG (red curve). The \(T_{av}\) is 1.00. The \(R_{FP}\) is 0.03. (c) The out-of-plane magnetic field of a Z-shaped \(S=-1\) IA-PhCWG at \(a/\lambda = 0.277\) with a transmittance of 0.97. (d) PBS of the \(S=-1\) IS-PhCWG. The black arrow shows the single-mode region. The inset shows one corner of the bent waveguide. (e) The transmittance spectra of the straight \(S=-1\) IS-PhCWG (black curve) and Z-shaped bent \(S=-1\) IS-PhCWG (red curve). The \(T_{av}\) is 1.00. The \(R_{FP}\) is 0.17.(f) The out-of-plane magnetic field of a Z-shaped \(S=-1\) IS-PhCWG at \(a/\lambda = 0.273\), with a transmittance of 0.96. }
\label{fig:figure5}
\end{figure}

\subsection*{(ii)S = -1, mirror-symmetric waveguides}
\(S=-1\) PhCWGs have another type of zigzag interface where the lattice configuration is more compact. Similar to that in the honeycomb lattice ones \cite{Mref26_h13_Kumar2022,Mref27_H13_Arora2021,Mref28_H13_Yang2020}, the waveguide mode has odd spatial parity about the central waveguide line. As shown in Fig.3(a), the \(S=-1\) IA-PhCWG has a wide single mode in \(a/\lambda = 0.263-0.292\). As shown in Fig.3(b), the transmittance is very high throughout the whole single-mode region with weak F-P ripples. The \(T_{av}\) is 1.00. The \(R_{FP}\) is 0.03. Figure3(c) shows \(H_z\) distribution at \(a/\lambda = 0.267\), where the transmittance is 0.90. The simulated results of \(S=-1\) IA-PhCWG agree well with a recent report by He et al.\cite{Mref4.1_HeXT2022_t13}.

S = -1 IS-PhCWG has a single-mode region in \(a/\lambda = 0.260-0.284\). As shown in Fig.3(e), the bend-transmittance fluctuates around unity with obvious F-P ripples in the spectrum. The cause of this irregular spectrum is the same as that for the \(S=1\) IS-PhCWG. The \(T_{av}\) of \(S=-1\) IS-PhCWG is 1.00, and the \(R_{FP}\) is 0.17. Figure5(f) shows \(H_z\) distribution at \(a/\lambda = 0.273\), where the transmittance is 0.96. Similar to the \(S=1\) and \(S=-2\) waveguides, the \(S=-1\) PhCWGs have high bend-transmission regardless of the inversion symmetry in the bulk lattice.

\subsection*{(iii)S = 2, glide-symmetric waveguides}
S=2 PhCWGs correspond to another type of bearded interface\cite{Mref10_Yoshimi:20_h56,Mref12_Gao2017_h56,Mref9_Yoshimi:21_h56,Mref25_Guillermo_disorderH23,Mref8_JalaliMehrabad2020_h56} other than S=-2 waveguides. It has been reported that the lower band of this domain-wall type in honeycomb-lattice VPhCs has high bend-transmission while the upper band has low bend-transmission \cite{Mref9_Yoshimi:21_h56,Mref10_Yoshimi:20_h56,Mref8_JalaliMehrabad2020_h56}. Here, in a triangular lattice PhCWG, the lower band has no single-mode region. Here, we focus only on the upper band. As shown in Fig.4(a,d), the single-mode region is 0.271-0.301 for IS-PhCWG and 0.277-0.306 for IA-PhCWG. As shown in Fig.4(b,e), both \(S=2\) PhCWGs have low bend-transmission in the upper band. The \(T_{av}\) is 0.33 for IA-PhCWG and 0.14 for IS-PhCWG, both lower than that of a W1WG. The calculated \(R_{FP}\) is 1.00 for both IA-PhCWG and IS-PhCWG. Figure4(c) shows the \(H_z\) distribution of \(S=2\) IA-PhCWG at \(a/\lambda = 0.285\), where the transmittance is 0.39. Figure4(f) shows the \(H_z\) distribution of \(S=2\) IS-PhCWG at \(a/\lambda = 0.279\), where the transmittance is 0.17. In Fig.4(c,f), there is an apparent intensity attenuation during propagation and strong backscattering at the left side of the wave sources.
\begin{figure}[hbt!]
\includegraphics[width=1.0\textwidth]{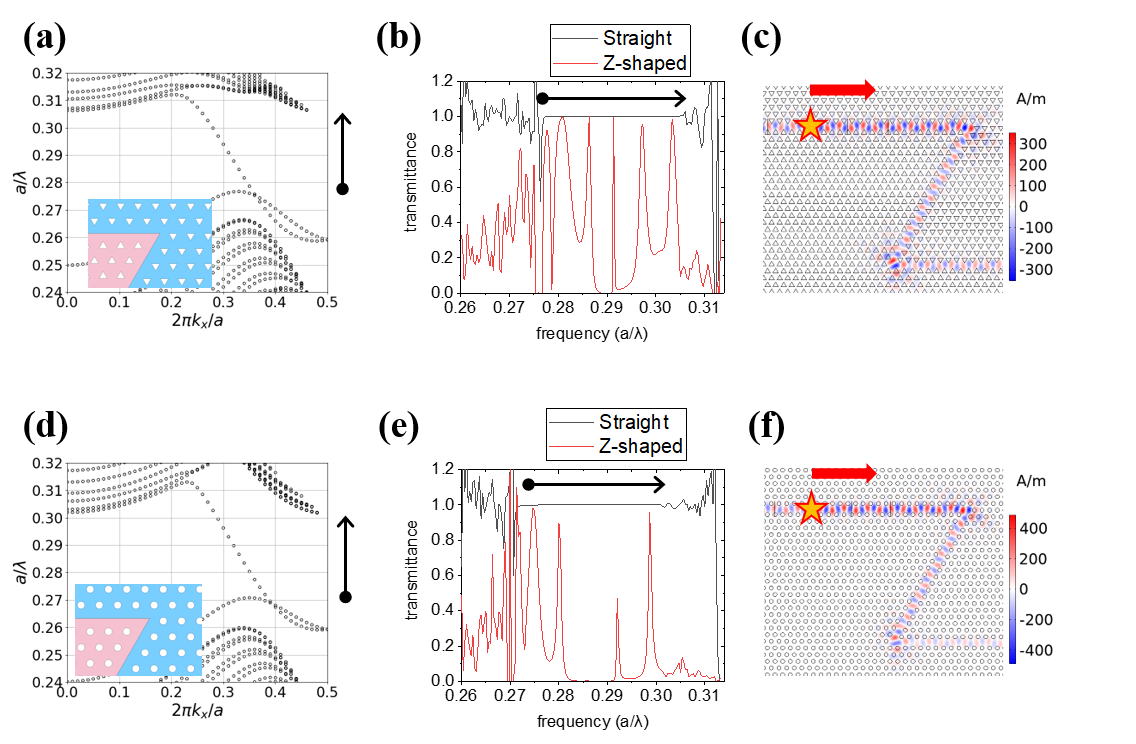}
\caption{Calculation results of the straight and Z-shaped \(S=2\) PhCWGs. (a) PBS of the \(S=2\) IS-PhCWG. The black arrow shows the single-mode region. The inset shows one corner of the bent waveguide. (b) The transmittance spectra of the straight \(S=2\) IS-PhCWG (black curve) and Z-shaped bent \(S=2\) IS-PhCWG (red curve). (c) The out-of-plane magnetic field of a Z-shaped \(S=2\) IS-PhCWG at \(a/\lambda = 0.279\) with a transmittance of 0.17. (d) PBS of the \(S=2\) IA-PhCWG. (e) The transmittance spectra of the straight \(S=2\) IA-PhCWG (black curve) and Z-shaped bent \(S=2\) IA-PhCWG (red curve). (f) The out-of-plane magnetic field of a Z-shaped \(S=2\) IA-PhCWG at \(a/\lambda = 0.285\) with a transmittance of 0.39. }
\label{fig:figure6}
\end{figure}
\clearpage

In this waveguide design, we are not able to evaluate the lower band of \(S=2\) PhCWGs. However, one can obtain single modes for both bands in an \(S=2\) PhCWG by locally changing the diameter of air holes along the waveguide interface. For a full comparison with the previous VPhCWGs, we discuss the hole-resized \(S=2\) PhCWGs in section 3. 

\subsection*{(iv)S = 3, mirror-symmetric waveguides without inversion symmetry}

Finally, we return to \(S=3\) PhCWG with triangular holes (broken inversion symmetry). Similar to \(S=3\) IS-PhCWG (W1WG), the lower band has a wide single-mode region. This interface does not correspond to any conventional VPhCWGs. As shown in Fig.5(a), the single-mode region is 0.270-0.293. As shown in Fig.5(b), the bend-transmission is very low. The \(T_{av}\) is 0.41. The \(R_{FP}\) is 0.56. Figure5(c) shows the \(H_z\) distribution of \(S=3\) IA-PhCWG at \(a/\lambda = 0.275\), where the transmittance is 0.58. There is an apparent attenuation and distortion of the \(H_z\)  in the propagation and strong backscattering at the left side of the wave source.

\begin{figure}[hbt!]
\includegraphics[width=1.0\textwidth]{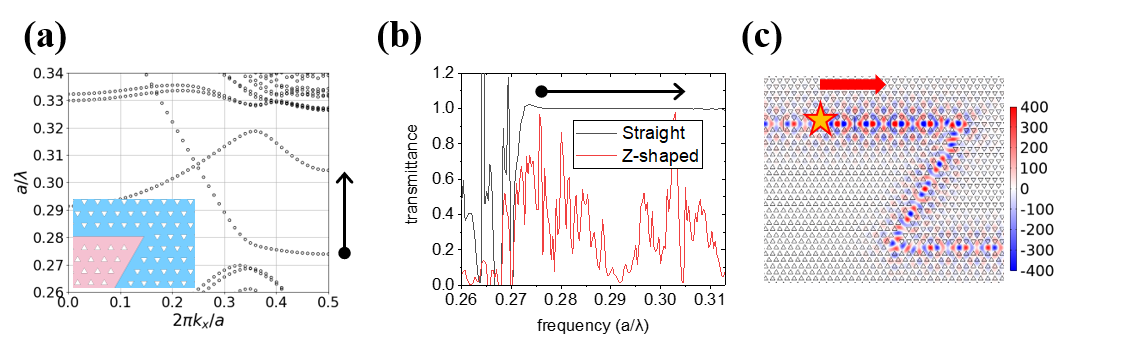}
\caption{Calculation results of the straight and Z-shaped \(S=3\) IA-PhCWGs. (a) PBS of the \(S=3\) IA-PhCWG. The black arrow shows the single-mode region. The inset shows one corner of the bent waveguide. (b) The transmittance spectra of the straight \(S=3\) IA-PhCWG (black curve) and Z-shaped bent \(S=3\) IA-PhCWG (red curve). The \(T_{av}\) is 0.41. The \(R_{FP}\) is 0.56. (c) The out-of-plane magnetic field of a Z-shaped \(S=3\) IA-PhCWG at \(a/\lambda = 0.275\) with a transmittance of 0.58.}
\label{fig:figure7}
\end{figure}

\section*{2. Excluding trivial factors}
We have discussed the influence of domain-wall types and inversion symmetry breaking on bend-transmission. In Table 1 we demonstrate that other factors such as spatial parity, the waveguide width, group refractive index (\(n_g\)), and band index do not affect the bend-transmission of waveguide modes. While it is known that bend scattering is strong in slow light regions, the focus of this work is the overall performance of waveguide bands in a broad wavelength range. The \(n_g\) here is calculated at the middle of the single-mode region for each waveguide band. We can see that the waveguide mode can have either high bend-transmittance or low bend-transmittance regardless of the value of \(n_g\). The same can be observed for other properties like spatial parity and band index.

\begin{table}[hbt!]
\includegraphics[width=0.9\textwidth]{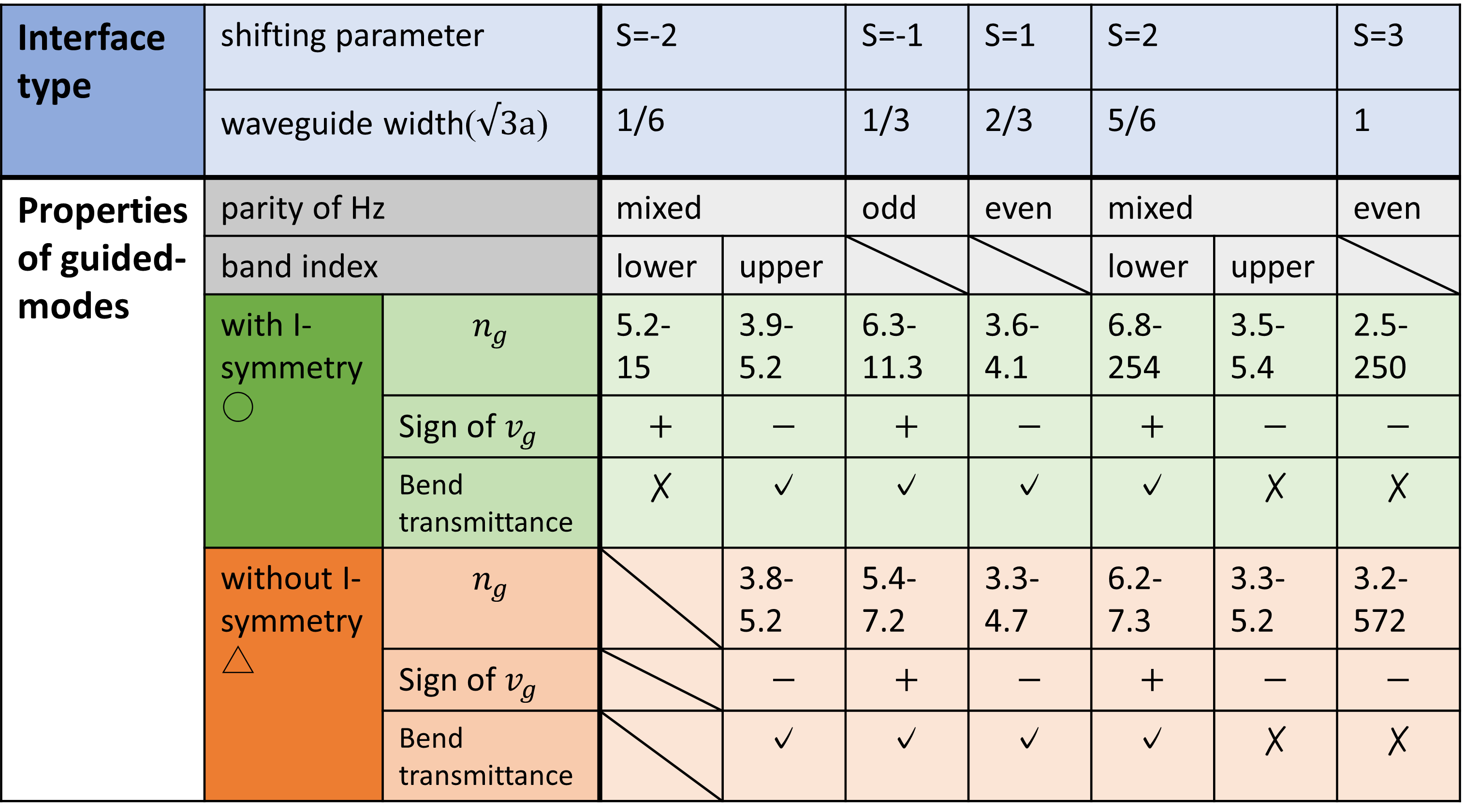}
\caption{Summary of properties of waveguide modes by their domain-wall types (S parameters) and bulk lattice symmetry. Check marks indicate high and "X" marks indicate low bend transmittance.}
\label{table S1}
\end{table}
\clearpage

\section*{3. Resizing air holes}
By locally resizing the air holes in the PhCWG, we can further engineer the band structure. Here we demonstrate 2 examples: the shrunken S=0 PhCWG and the expanded \(S=2\) PhCWG.
\subsection*{i)	The shrunken S=0 PhCWG}	
Without lattice shifting, we directly introduce a defect to the bulk triangular lattice by shrinking 2 arrays of air holes along the \(\Gamma K\) direction, as shown in the inset image Fig.6(a).  In the simulation, the lattice constant is 500 nm. The larger air holes in the bulk lattice have a radius of 136 nm, and the smaller air holes at the interface have a radius of 81 nm. The effective refractive index is 2.74.  The shrunken S=0 PhCWGs have glide-symmetry along the interface. As shown in Fig.6(a), the band structure is more complicated than previously discussed waveguides, with 4 bands and 2 degeneracies in the bandgap. Here only one band has single modes at \(a/\lambda=0.309-0.315\). The bend-transmittance of this band is very high as shown with red curves in Fig.6(b). The \(T_{av}\) is 0.933. The \(R_{FP}\) is 0.149. We can also confirm from the \(H_{z}\) profile in Fig.6(c) that the bend-transmission is high with no apparent backward flux at the left side of the wave source (five-point star).

\begin{figure}[hbt!]
\includegraphics[width=1.0\textwidth]{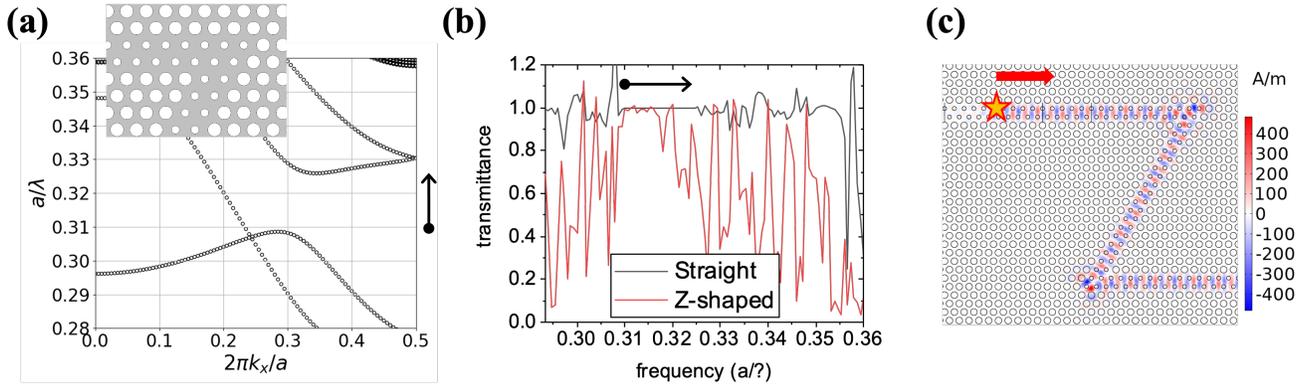}
\caption{Calculation results of the straight and Z-shaped \(S\) = 0 IS-PhCWGs. (a) PBS of the \(S = 0\) Cir-PhCWG. The black arrow shows the single-mode region. The inset shows one corner of the bent waveguide. The larger air holes in the bulk lattice have a radius of 136 nm. The small air holes have a radius of 81 nm. (b) The transmittance spectra of the straight \(S = 0\) IS-PhCWG (black curve) and Z-shaped bent \(S = 0\) IS-PhCWG (red curve). (c) The out-of-plane magnetic field of a Z-shaped \(S = 0\) IS-PhCWG at \(a/\lambda=0.313\) with a transmittance of 1.00.}
\label{fig:figure.S2}
\end{figure}

% We have further fabricated and measured the S=0 IS-PhCWG. The 3-dimensional waveguide has 3 degeneracies at the BZ edge. Due to fabrication imperfections, not all of these single modes are observed in our measurement. As shown in Fig.2 (c), the bend-transmission in the wavelength range 1360 nm-1400nm is (red for evenly bent and blue for unevenly bent waveguides) comparable to that of the straight waveguide (black curve). And the F-P ripples are almost invisible, demonstrating extreme robustness considering there are 7 bends. At wavelength 1400 nm-1450nm, the bend-transmittance drops drastically, corresponding to the low bend-transmission band with lower frequency, which cannot be evaluated in the simulation. This result serves as a clear demonstration that robust propagation via 120-degree sharp bends does not require inversion-symmetry breaking, and thus does not involve valley-photonics. Regarding the mode classification of the bands in S=0 IS-PhCWG, we speculate that they are related to an S=6 PhCWG, which has two arrays of air holes removed along the GK direction. Due to the complicated band evolution as S grows larger than 3, we do not have a clear answer currently. 

\subsection*{ii)	The expanded \(S=2\) PhCWG}
The expanded \(S=2\) PhCWGs have the same lattice configuration as the \(S=2\) PhCWGs except that the nearest 2 arrays of air holes along the interface have larger sizes than those in the bulk lattice. The \(S=2\) IS-PhCWGs have a lattice constant of 460 nm. The radius of larger holes is 204 nm. The radius of the smaller holes is 146 nm. The effective refractive index of silicon is set to be 2.7. As the hole size grows larger along the interface, the lower of the two touching bands rises up and forms a single-mode region, making it possible to investigate its transmission property (Fig.7(a)). As shown in Figure7(b) the lower band has high transmission through Z-shaped waveguides.  The \(T_{av}\) is 1.00 in the upper band and 0.13 in the lower band. The \(R_{FP}\) is 0.18 in the lower band and 1.00 in the upper band.

\begin{figure}[hbt!]
\includegraphics[width=1.0\textwidth]{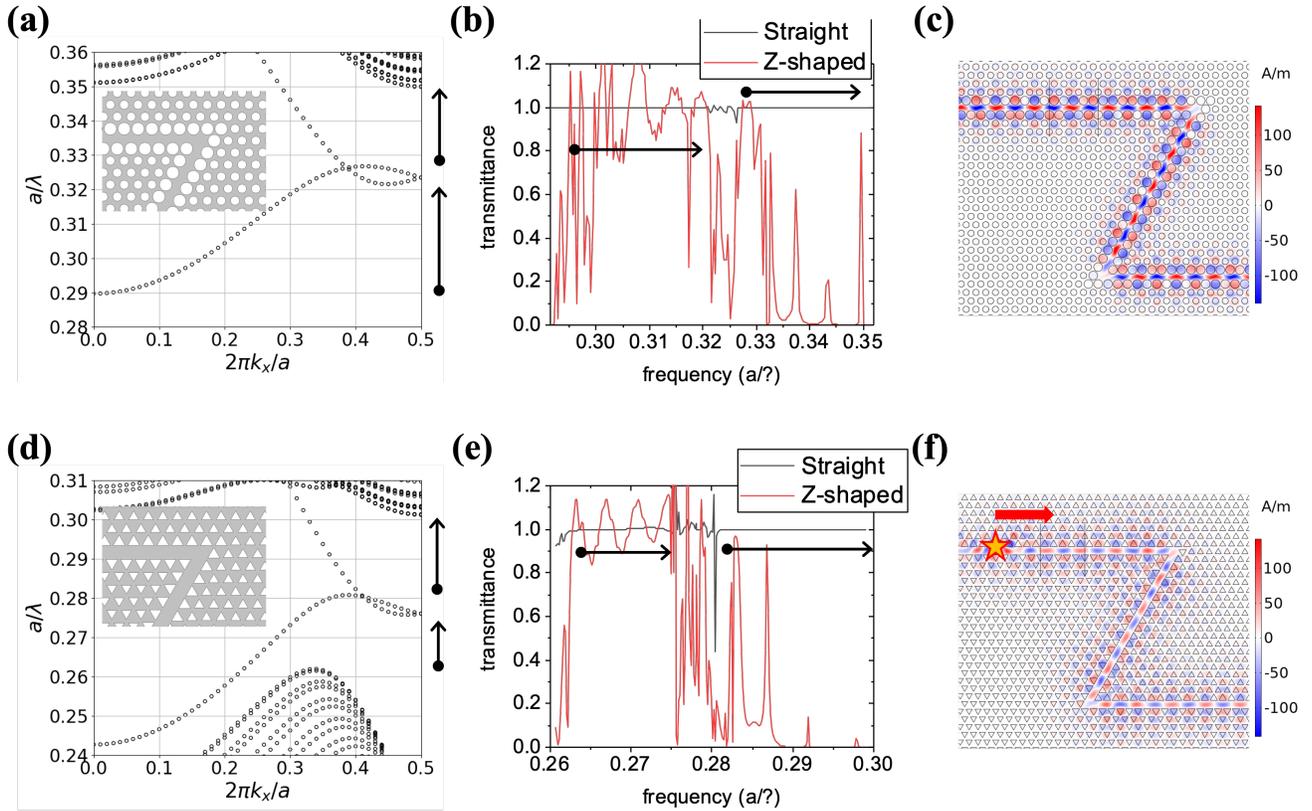}
\caption{Calculation results of the straight and Z-shaped \(S=2\) PhCWGs. (a) PBS of the \(S=2\) IS-PhCWG. The black arrow shows the single-mode region. The inset shows one corner of the bent waveguide. The large air holes in the bulk lattice have a radius of 204 nm. The small holes have a radius of 146 nm. (b) The transmittance spectra of the straight \(S=2\) IS-PhCWG (black curve) and Z-shaped bent \(S=2\) IS-PhCWG (red curve). (c) The out-of-plane magnetic field of a Z-shaped \(S=2\) IS-PhCWG at \(a/\lambda=0.313\) with a transmittance of 0.96. (d) PBS of the \(S=2\) IA-PhCWG. The black arrow shows the single-mode region. The inset shows one corner of the bent waveguide. The large air holes in the bulk lattice have a side length of 381 nm. The small holes have a side length of 319 nm. (e) The transmittance spectra of the straight \(S=2\) IA-PhCWG (black curve) and Z-shaped bent \(S=2\) IA-PhCWG (red curve). (f) The out-of-plane magnetic field of a Z-shaped \(S=2\) IS-PhCWG at \(a/\lambda=0.267\) with transmittance of 1.}
\label{fig:figure.S3}
\end{figure}
\clearpage

The \(S=2\) IA-PhCWGs have a lattice constant of 460 nm. The side length of the larger triangular holes is 382 nm. The side length of the smaller triangular holes is 319 nm. The effective refractive index of silicon is 2.7. Like the IS-PhCWGs, the lower band of IA-PhCWGs rises up due to the resizing of air holes, as shown in Fig.7(d). The \(T_{av}\) is 1.00 in the upper band and 0.05 in the lower band. The \(R_{FP}\) is 0.07 in the lower band and 1.00 in the upper band. 

We have included the results of the hole-resized \(S=2\) PhCWGs in Table.1 and Fig.7 in the main manuscript.

\section*{4. Experiment setting}

We conduct the transmission measurement of waveguides using the experimental setup shown in Fig.8. Light from a wavelength-tunable laser is collimated by lens 1 and obtains linear polarization through POL 1. The polarization direction can be adjusted by rotating the half-wave plate (HWP) 1. In our measurement, we set the orientation of the HWPs so that only TE waves are transmitted into the device. Then the incident light is focused by lens 2 and coupled to the waveguide. The transmitted light undergoes a reversed process and is collected by the power meter. The power meter measures light intensity in decibel-milliwatts. The measurable power is up to -110dBm. There are time-variant fluctuations in the measured intensity up to 1dBm. The measured spectra are smoothed using adjacent averaging to remove the irrelevant fluctuation in raw data. 
\begin{figure}[hbt!]
\includegraphics[width=0.9\textwidth]{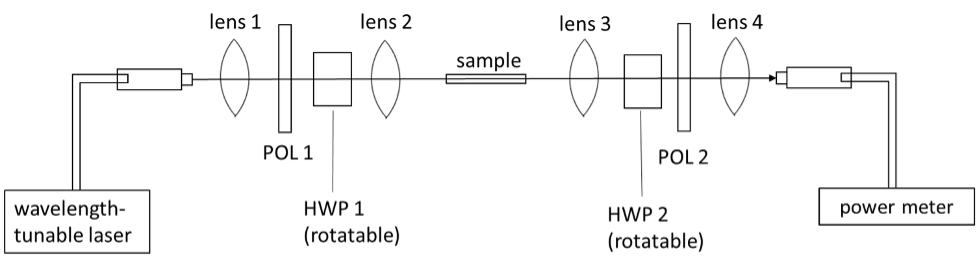}
\caption{Schematic of the experimental setup. Light from a wavelength-tunable laser is collimated by lens 1 and obtains linear polarization through POL 1. The polarization direction can be adjusted by rotating the half-wave plate (HWP) 1.}
\label{fig:figure.S4}
\end{figure}

\section*{5. Mode classification based on band structures}

We have discussed various waveguides with integer \(S\) values. Actually, the \(S\) can be tuned continuously. We calculate the photonic band structure for various \(S\) values in a 3-dimensional Si PhC slab in Fig.9. As the shifting parameter \(S\) decreases continuously, the waveguide modes emerge from bulk modes below the PBG and move upwardly into the bulk modes above the PBG. Different bands can be traced to one another in this process. Thus, we can classify these bands into four groups as \(S\) changes from 3 to -2.  The even mode of W1WG and the upper band of \(S=2\) glide-symmetric PhCWG belong to group 1 (green). The lower band of \(S=2\) glide-symmetric PhCWG and the even band of \(S=1\) PhCWG belong to group 2 (blue). The odd mode of \(S=-1\) PhCWG and the upper band of \(S=-2\) glide-symmetric PhCWG belong to group 3 (orange). Finally, the lower band of \(S=-2\) glide-symmetric PhCWG belongs to group 4 (gray). Waveguide modes in group 1 and group 4 have low bend-transmission, and those in group 2 and group 3 have high bend-transmission. This mode classification originates from the photonic band structures (PBGs). Therefore, it is not self-evident that the bend-transmission via 120-degree bends is related to this mode classification.

\begin{figure}[hbt!]
\includegraphics[width=1.0\textwidth]{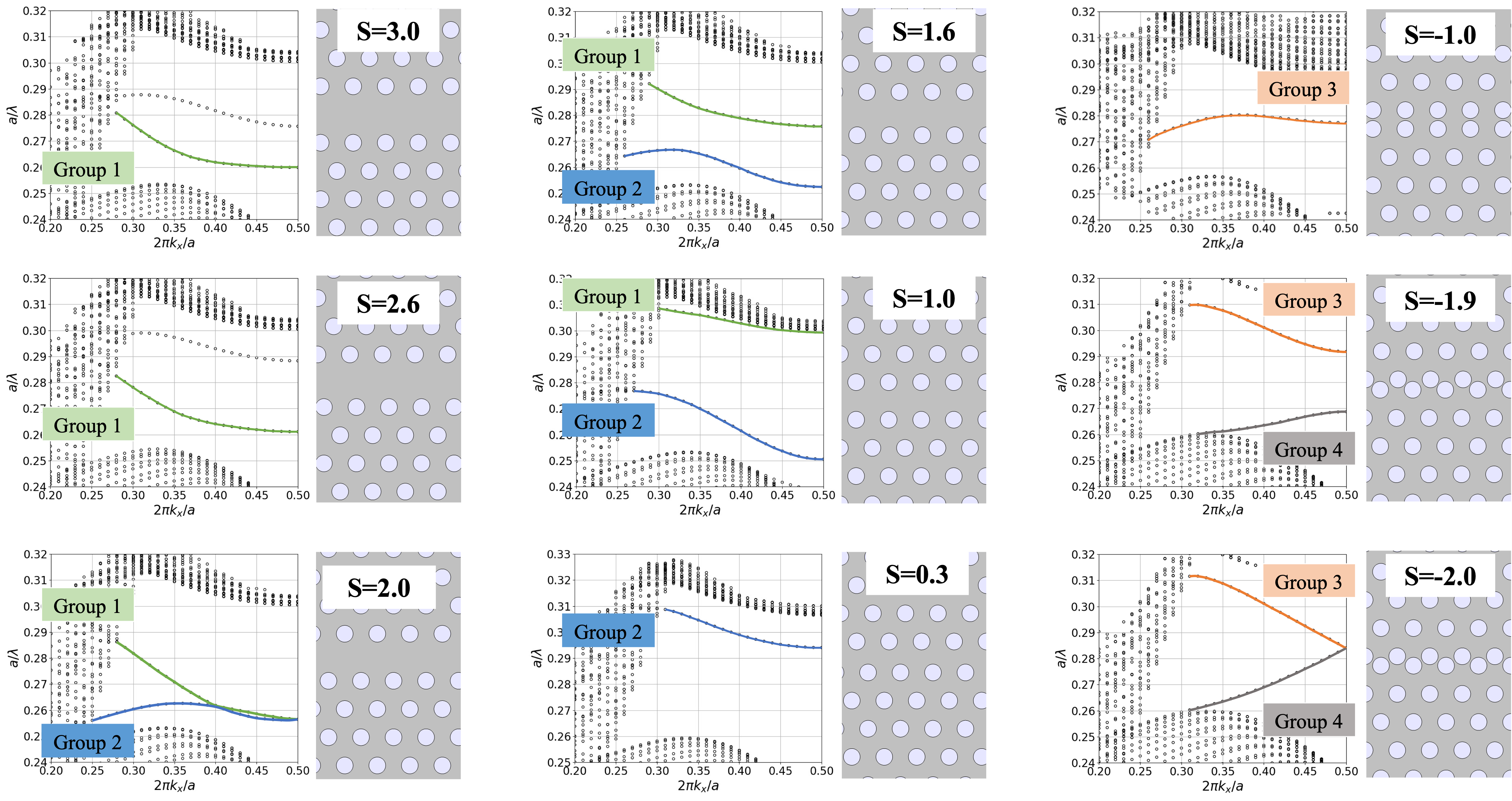}
\caption{Band structure of PhCWGs of various domain-wall types as shifting parameter \(S\) changes continuously from 3 to -2. Bands of the same group can be traced to each other. }
\label{fig:figure10}
\end{figure}
\clearpage

\section*{6. Simulation results of circular polarization singularities}
We have investigated the spatial distribution of circular polarization singularities or C-points (CPs) for all domain-wall types discussed in the main text. Figure10 shows the CP distribution inside an \(S=-2\) IA-PhCWG in the high bend-transmission band. The top four images show the zero-value isolines of the Stokes parameters, where the crossing node of the \(S_1=0\) and \(S_2=0\) lines represent the location of CPs. The bright CPs near the connection interface are marked with black boxes. The bottom images show the normalized E-field intensities at the corresponding frequencies. Below each image are the normalized frequencies (\(a/\lambda\)), normalized intensities at the location of the CPs, and the bend-transmittance of the Z-shaped waveguide. For waveguide modes at \(a/\lambda=0.269\) and \(a/\lambda=0.282\), there are bright CPs inside the first nearest air holes. At higher frequencies, the CPs disappear from these holes although the degree of polarization is still very high (over 0.99). New CPs appear inside the second nearest air holes. This change of locations differs from the case in \(S=-2\) IS-PhCWGs, which indicates that the shape of air holes can modify the location of CPs to some extent. 
\begin{figure}[hbt!]
\includegraphics[width=0.9\textwidth]{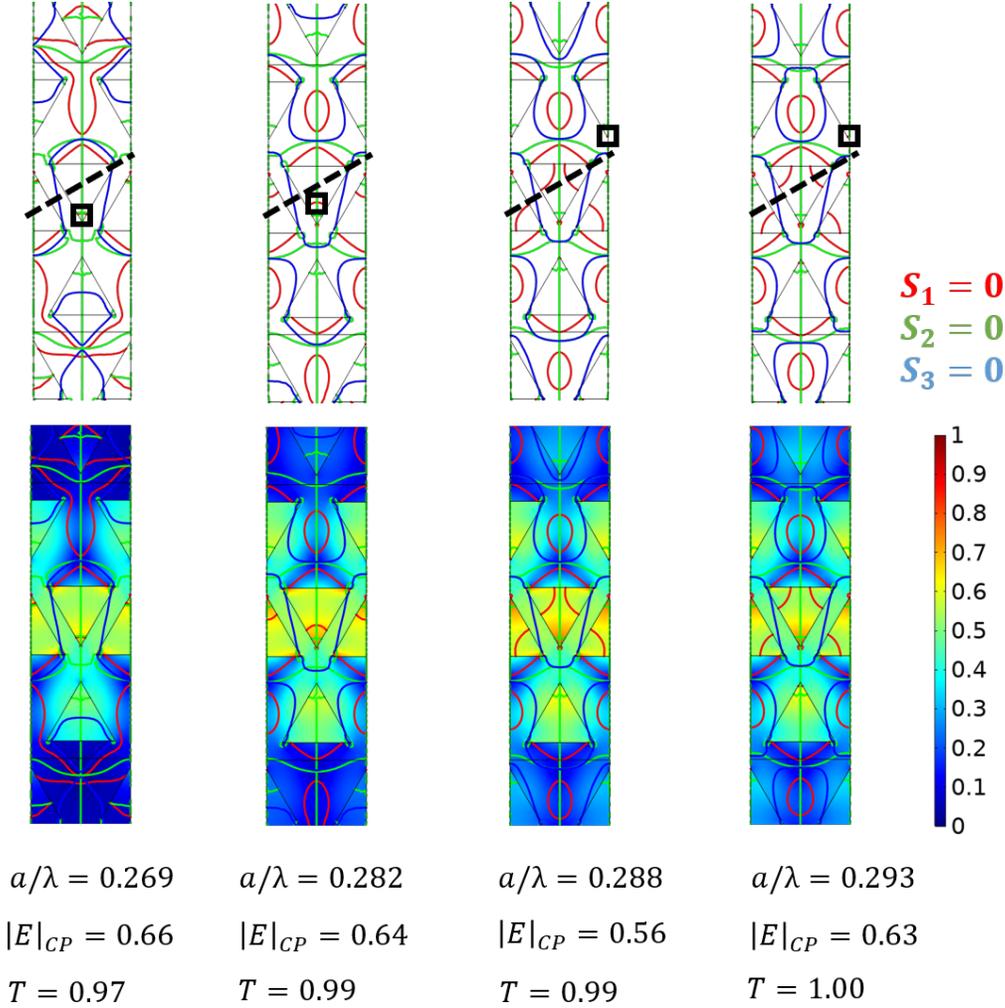}
\caption{The zero-value isoline of Stokes parameters (top) and the Electric field amplitudes (bottom) of \(S=-2\) IA-PhCWGs. The corresponding frequencies, normalized E-field amplitudes at the location of CPs, and the bend-transmittance are shown below each plot.}
\label{fig:figure.S5}
\end{figure}
\begin{figure}[hbt!]
\includegraphics[width=1.0\textwidth]{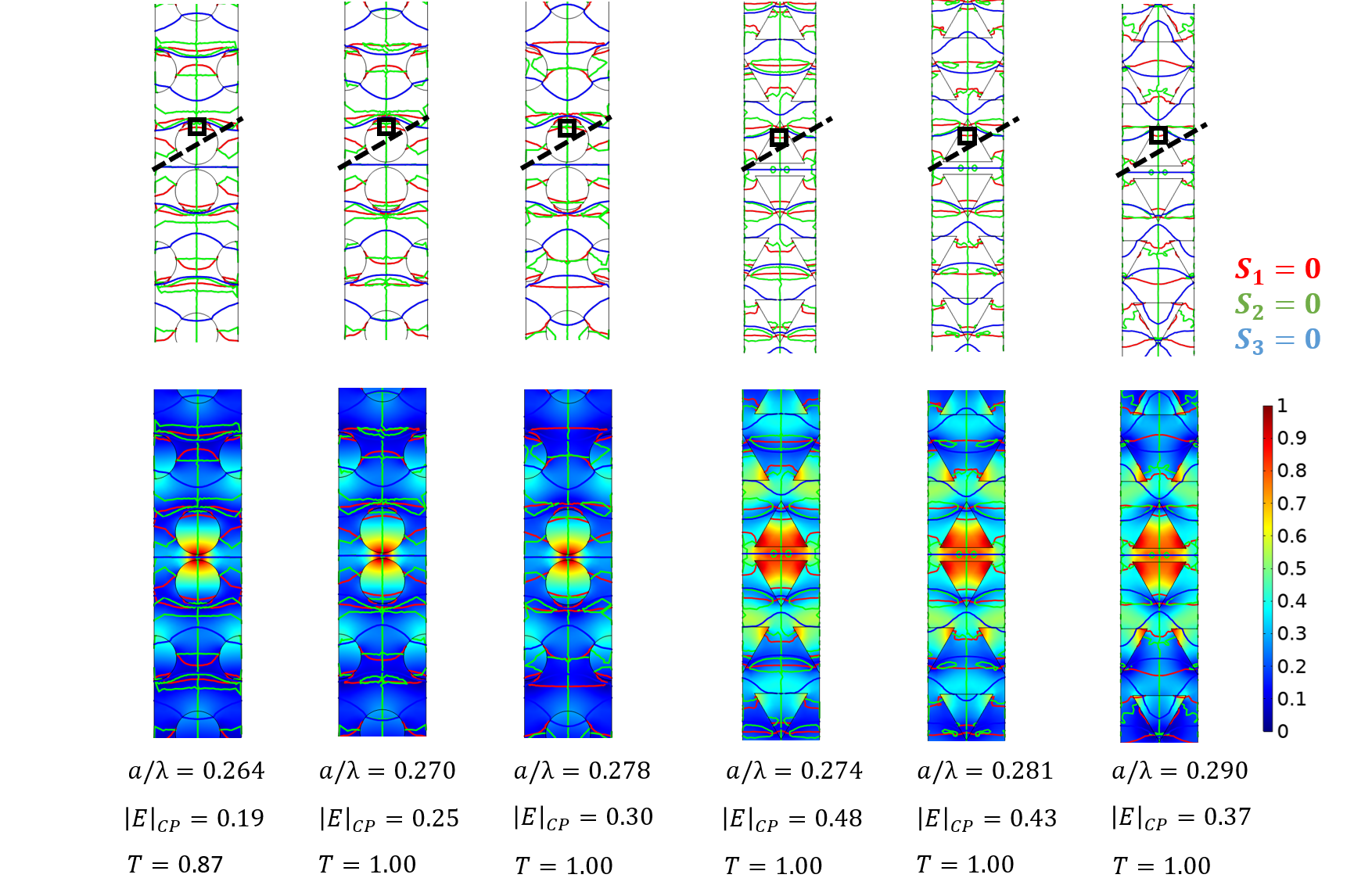}
\caption{The zero-value isoline of Stokes parameters (top) and the Electric field amplitudes (bottom) of \(S=-1\) IS- (left three columns) and IA-(right three columns) PhCWGs. The corresponding frequencies, normalized E-field amplitudes at the location of CPs, and the bend-transmittance are shown below each plot.}
\label{fig:figure.S6}
\end{figure}

Figure11 shows the CP distributions inside \(S=-1\) PhCWGs. The left six images show the \(S=-1\) IS-PhCWGs and the right six images show the \(S=-1\) IA-PhCWGs. In both cases, there are bright CPs inside the nearest air holes near the connection interfaces. Figure12 shows the case of \(S=1\) PhCWGs. The left two images show the result of \(S=1\) IS-PhCWGs. We plot the result of only one frequency for the \(S=1\) IS-PhCWG because the CPs and field intensities do not change much within the single-mode region due to the narrow waveguide band. The right six images show the result of \(S=1\) IA-PhCWGs. In the \(S=1\) PhCWGs, the bright CPs near the connection interface are located inside the silicon area instead of inside the air holes. We speculate that CPs in the silicon areas are more susceptible to perturbations and less stable at the bends, which explains the relatively larger fluctuations in the transmittance spectra observed in the \(S=1\) PhCWGs. Actually, when we restore the second sublattice in the \(S=1\) PhCWGs (a zigzag interface PhCWG having a honeycomb lattice), there are also bright CPs at similar locations, except that these CPs fall within the air holes of the second sublattice. We have found that the simulated transmittance spectrum of the bent honeycomb lattice \(S=1\) PhCWG is more stable than its triangular lattice counterparts.
\begin{figure}[]
\includegraphics[width=1.0\textwidth]{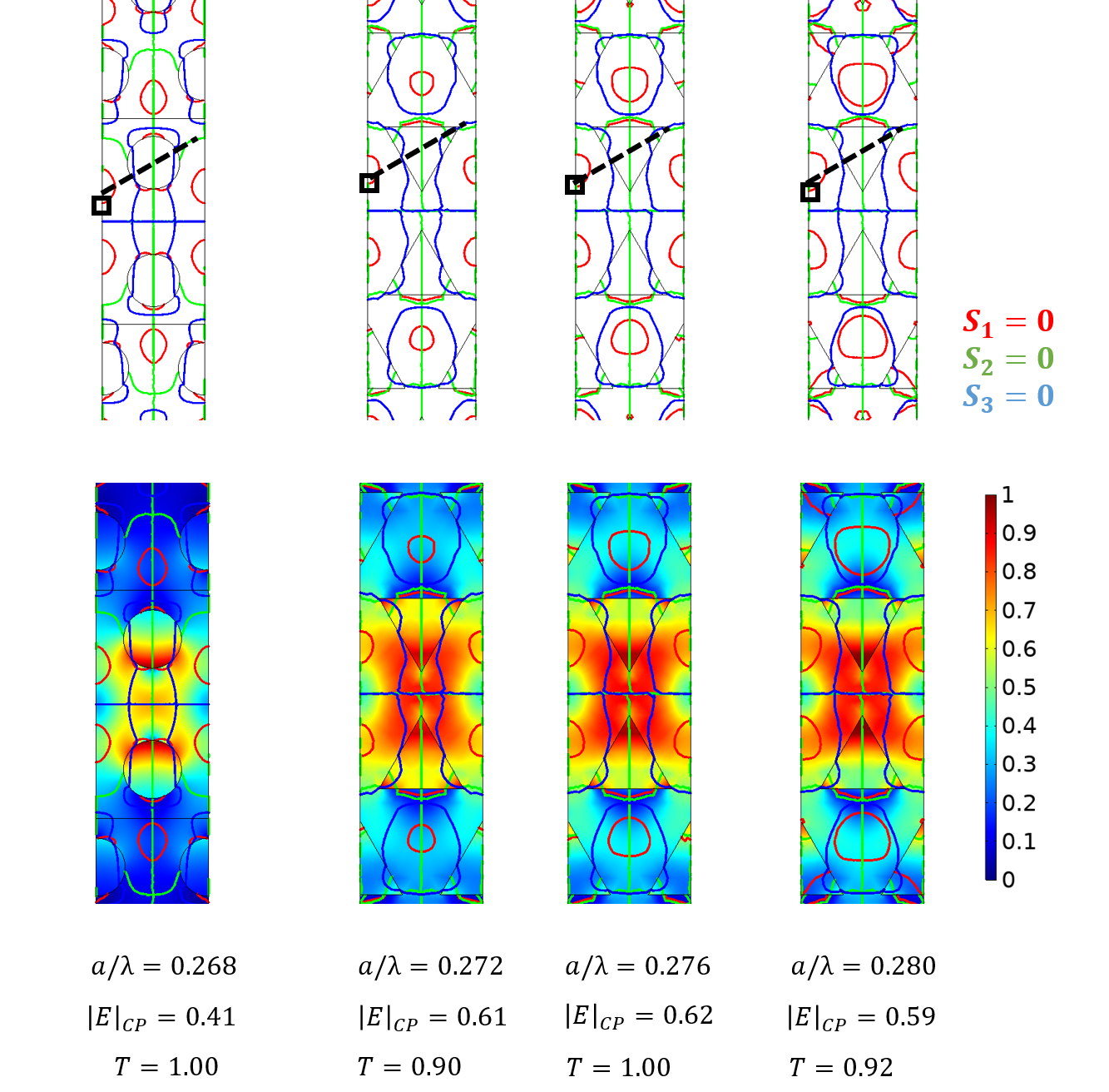}
\caption{The zero-value isoline of Stokes parameters (top) and the Electric field amplitudes (bottom) of \(S=1\) IS-(left one column) and IA-(right three columns) PhCWGs. The corresponding frequencies, normalized E-field amplitudes at the location of CPs, and the bend-transmittance are shown below each plot.}
\label{fig:figure.S7}
\end{figure}

Figure13 shows the results of the \(S=2\) IS-PhCWGs. If we ignore the "unstable" CPs inside the silicon areas,  we can see that the high bend-transmission band has bright CPs near the connection interface, which lies inside the air holes, while in the low bend-transmission band these bright CPs disappear. We apply the same rule to the S-3 PhCWGs and ignore the silicon region CPs. The left six images in Fig.14 show the results of \(S=3\) IS-PhCWGs. For lower frequencies such as \(a/\lambda=0.269\), there are bright CPs near the connection interface. However, due to the small group velocity and F-P resonance, the bend-transmittances are very low. At higher frequencies, the bright CPs inside the air holes disappear, thus no stable, bright CPs near the connection interface. The right six images in Fig.14 show the results of \(S=3\) IA-PhCWGs. For the \(S=3\) IA-PhCWGs, we find that at higher frequencies such as  \(a/\lambda=0.272\) and \(a/\lambda=0.276\) there exist "stable", bright CPs near the connection interface. Interestingly, the bend-transmittances near these frequencies are relatively higher. We have observed that the \(S=3\) IA-PhCWG has higher bend-transmittance than the \(S=3\) IS-PhCWG overall. Maybe this is related to the "stable", bright CPs we have found in the \(S=3\) IA-PhCWGs. In addition, this is another case where the shapes of air holes mildly modify the location of CPs and affect the bend-transmittance when the domain-wall type is not changed.

To conclude our preliminary investigation into the circular polarization singularities, we have observed the existence of air-hole bright CPs in several high bend-transmittance bands and their disappearance in some low bend-transmittance bands. However, in \(S=2\) and \(S=3\) PhCWGs, there are bright CPs in the silicon region near the connection interface. We speculate that multiple CPs can appear inside the dielectric region with strong field intensity when the waveguide width is large. However, because they are exposed to a large area of uniform dielectric, these dielectric CPs can be unstable at the bends where the translational symmetry is broken. Furthermore, because the lattice configuration is more compact at the bends in narrower PhCWGs, CPs are less affected by the broken translational symmetry. The most important point is that CPs located inside the air holes disappear around the position where the bend-transmission abruptly decreases. This suggests that there is some correlation between the high bend-transmission and the existence of CPs. 

Finally, we give an example that the bright CPs continue to exist inside the bending corners of a Z-shaped waveguide (Fig.15). In the high bend-transmittance band of \(S=-2\) IA-PhCWG, we have confirmed the existence of bright CPs near the connection interface in the unit cell calculation. Here we investigate whether the same CPs persist in the bends if we excite the same waveguide mode in a Z-shaped waveguide. Like in the unit cell calculations, we identify the locations of CPs using the Stokes parameter isolines. The bright CPs near the bending corners are marked with black boxes. In the straight waveguide segment, we can find the bright CPs at the same location as those in the unit cell calculation. In the bends, we can see that there are strong distortions in the Stokes parameter isolines but the CPs (crossing nodes of red and green lines) exist in most of the air holes. After the light wave propagates through the second bend, the Stoke parameter isolines restore their shapes and the CPs are at the same location as before. While it is possible that the perturbation at the bends may give rise to accidental CPs. We believe this result demonstrates an uninterrupted distribution of the same type of CPs as we have identified in the unit cell calculation.
\begin{figure}[]
\includegraphics[width=1.0\textwidth]{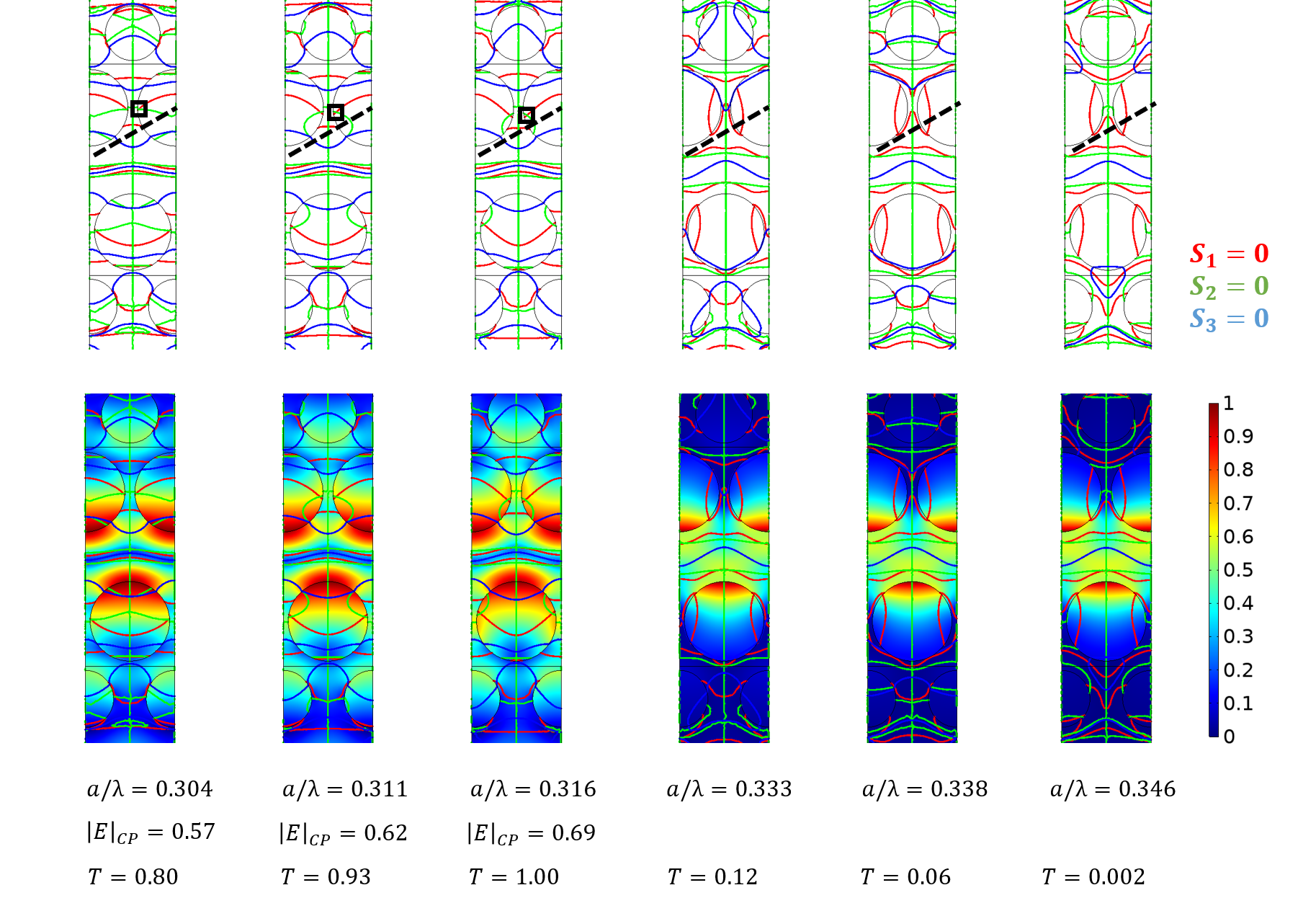}
\caption{The zero-value isoline of Stokes parameters (top) and the Electric field amplitudes (bottom) of \(S=2\) IS-PhCWGs. The corresponding frequencies, normalized E-field amplitudes at the location of CPs, and the bend-transmittance are shown below each plot.}
\label{fig:figure.8}
\end{figure}

\begin{figure}[]
\includegraphics[width=1.0\textwidth]{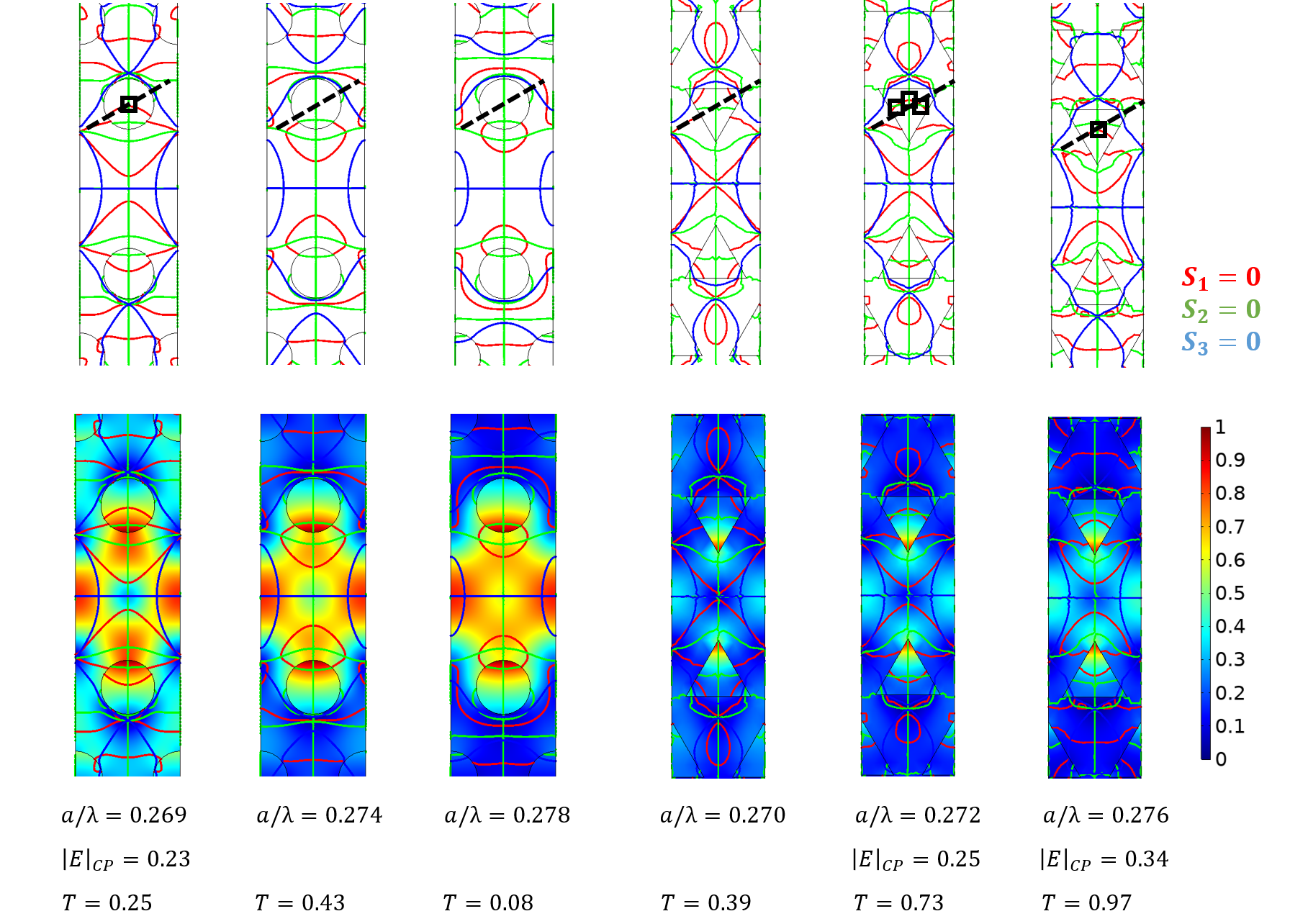}
\caption{The zero-value isoline of Stokes parameters (top) and the Electric field amplitudes (bottom) of \(S=3\) IS-(left three columns) and IA-(right three columns) PhCWGs. The corresponding frequencies, normalized E-field amplitudes at the location of CPs, and the bend-transmittance are shown below each plot.}
\end{figure}

\begin{figure}[]
\includegraphics[width=1.0\textwidth]{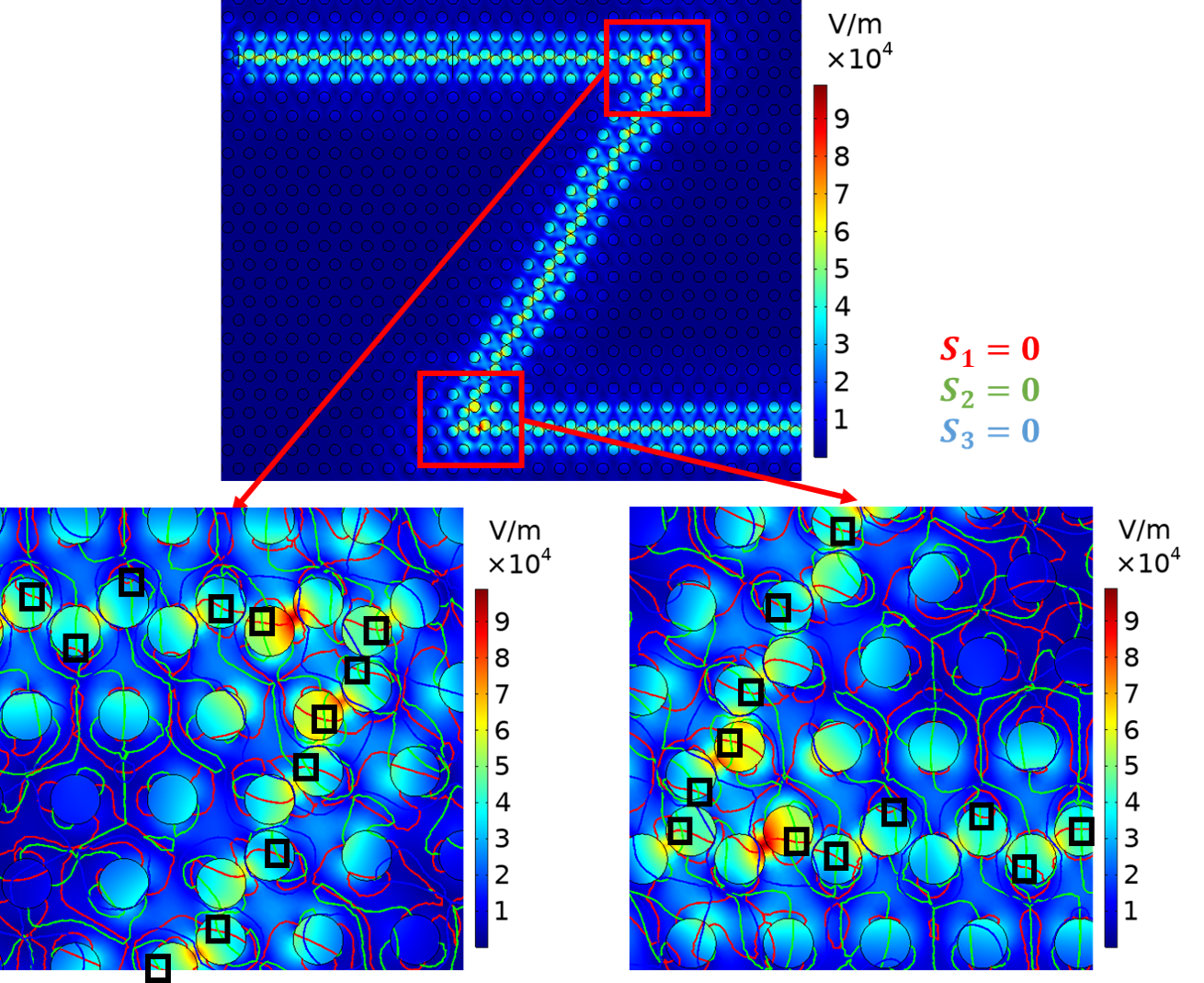}
\caption{The electric filed amplitude and CP distribution of a Z-shaped \(S=-2\) IS-PhCWG at frequency \(a/\lambda=0.291\). The locations of CPs are marked with black boxes.}
\label{fig:figure.S11}
\end{figure}

\bibliography{SI-ref}